\documentclass[12pt, letter]{article}
\usepackage[T1]{fontenc}
\usepackage[utf8]{inputenc}
\usepackage{amsmath,subfigure,  amsfonts, bm}
\newcommand{\RN}[1]{%
  \textup{\uppercase\expandafter{\romannumeral#1}}%
}
\usepackage{braket}
\usepackage{graphicx,epstopdf}
 \DeclareGraphicsExtensions{.png,.pdf}
\usepackage{natbib}
\usepackage{indentfirst}
\usepackage{mathabx}
\usepackage{wrapfig}
\usepackage{caption}
\usepackage{tabularx}
\usepackage{setspace}
\usepackage[toc,page]{appendix}
\usepackage{float}
\usepackage{booktabs}
\usepackage[margin = 1 in]{geometry}
\usepackage{amsmath,amsfonts,color,subfigure,amssymb,amsthm}
\usepackage[colorlinks=true,urlcolor=blue,citecolor=blue,linkcolor=blue,bookmarks=true]{hyperref}
\usepackage{xcolor}

\usepackage{algorithm}
\usepackage{algpseudocode}

\AtBeginDocument{}
\begin{document}

\title{Best Subset Selection: Statistical Computing Meets Quantum Computing}

\author{ }
\author{Wenxuan Zhong, Yuan Ke, Ye Wang, Yongkai Chen, Jinyang Chen, Ping Ma\thanks{Department of Statistics,  University of Georgia, Athens, GA 30602. E-mail:
\href{mailto:wenxuan@uga.edu}{\textsf{wenxuan@uga.edu, yuan.ke@uga.edu, ywang927@uga.edu, YongkaiChen@uga.edu, jingyang.chen@uga.edu, pingma@uga.edu}}.}
}
\date{}
\maketitle
\vspace{-0.5in}


\doublespacing



\newcommand{\stz}{|0\rangle}
\newcommand{\sto}{|1\rangle}
\newcommand{\sti}{|i\rangle}
\newcommand{\stbb}{|b\rangle}
\newcommand{\stistar}{|i^*\rangle}
\newcommand{\stik}{|i_k\rangle}
\newcommand{\stphi}{|\phi\rangle}
\newcommand{\stpsi}{|\psi\rangle}
\newcommand{\sta}{|a\rangle}
\newcommand{\stb}{|b\rangle}
\newcommand{\stalpha}{|\alpha\rangle}
\newcommand{\stbeta}{|\beta\rangle}

\newcommand{\braa}{\langle a|}

\newcommand{\xx}{\mathbf x}
\newcommand{\yy}{\mathbf y}
\newcommand{\bb}{\boldsymbol{\beta}}
\newcommand{\bmu}{\boldsymbol{\mu}}
\newcommand{\btheta}{\boldsymbol \theta}
\newcommand{\beps}{\boldsymbol{\epsilon}}
\newcommand{\sxx}{\mathbf{x}^*}
\newcommand{\sy}{y^*}
\newcommand{\sprob}{\pi^*}
\newcommand{\stheta}{\breve{\boldsymbol \theta}}
\newcommand{\htheta}{\hat{\btheta}}
\newcommand{\half}{\frac{1}{2}}
\newcommand{\mI}{\textbf I}
\newcommand{\dl}{ \dot{\ell} }
\newcommand{\ddl}{ \ddot{\ell} }
\newcommand{\stistark}{|i_k^*\rangle}
\newcommand{\stphiX}{|\psi_{\mathbf{X}}\rangle}
\newcommand{\stphiy}{|\psi_{\mathbf{y}}\rangle}
\newcommand{\stphixtilde}{|\psi_{\Tilde{\mathbf{x}}}\rangle}
\newcommand{\red}{\textcolor{red}}

\newcommand{\ab}{\mathbf{a}}
\newcommand{\bbb}{\mathbf{b}}
\newcommand{\cbb}{\mathbf{c}}
\newcommand{\db}{\mathbf{d}}
\newcommand{\eb}{\mathbf{e}}
\newcommand{\fb}{\mathbf{f}}
\newcommand{\gb}{\mathbf{g}}
\newcommand{\hb}{\mathbf{h}}
\newcommand{\ib}{\mathbf{i}}
\newcommand{\jb}{\mathbf{j}}
\newcommand{\kb}{\mathbf{k}}
\newcommand{\lb}{\mathbf{l}}
\newcommand{\mb}{\mathbf{m}}
\newcommand{\nbb}{\mathbf{n}}
\newcommand{\ob}{\mathbf{o}}
\newcommand{\pb}{\mathbf{p}}
\newcommand{\qb}{\mathbf{q}}
\newcommand{\rb}{\mathbf{r}}
\newcommand{\sbb}{\mathbf{s}}
\newcommand{\tb}{\mathbf{t}}
\newcommand{\ub}{\mathbf{u}}
\newcommand{\vb}{\mathbf{v}}
\newcommand{\wb}{\mathbf{w}}
\newcommand{\xb}{\mathbf{x}}
\newcommand{\yb}{\mathbf{y}}
\newcommand{\zb}{\mathbf{z}}

\newcommand{\bc}{\bm{c}}
\newcommand{\bd}{\bm{d}}
\newcommand{\be}{\bm{e}}
\newcommand{\bbf}{\bm{f}}
\newcommand{\bg}{\bm{g}}
\newcommand{\bh}{\bm{h}}
\newcommand{\bi}{\bmf{i}}
\newcommand{\bj}{\bm{j}}
\newcommand{\bk}{\bm{k}}
\newcommand{\bl}{\bm{l}}
\newcommand{\bbm}{\bm{m}}
\newcommand{\bn}{\bm{n}}
\newcommand{\bo}{\bm{o}}
\newcommand{\bp}{\bm{p}}
\newcommand{\bq}{\bm{q}}
\newcommand{\br}{\bm{r}}
\newcommand{\bs}{\bm{s}}
\newcommand{\bt}{\bm{t}}
\newcommand{\bu}{\bm{u}}
\newcommand{\bv}{\bm{v}}
\newcommand{\bw}{\bm{w}}
\newcommand{\bx}{\bm{x}}
\newcommand{\by}{\bm{y}}
\newcommand{\bz}{\bm{z}}

\newcommand{\Ab}{\mathbf{A}}
\newcommand{\Bb}{\mathbf{B}}
\newcommand{\Cb}{\mathbf{C}}
\newcommand{\Db}{\mathbf{D}}
\newcommand{\Eb}{\mathbf{E}}
\newcommand{\Fb}{\mathbf{F}}
\newcommand{\Gb}{\mathbf{G}}
\newcommand{\Hb}{\mathbf{H}}
\newcommand{\Ib}{\mathbf{I}}
\newcommand{\Jb}{\mathbf{J}}
\newcommand{\Kb}{\mathbf{K}}
\newcommand{\Lb}{\mathbf{L}}
\newcommand{\Mb}{\mathbf{M}}
\newcommand{\Nb}{\mathbf{N}}
\newcommand{\Ob}{\mathbf{O}}
\newcommand{\Pb}{\mathbf{P}}
\newcommand{\Qb}{\mathbf{Q}}
\newcommand{\Rb}{\mathbf{R}}
\newcommand{\Sbb}{\mathbf{S}}
\newcommand{\Tb}{\mathbf{T}}
\newcommand{\Ub}{\mathbf{U}}
\newcommand{\Vb}{\mathbf{V}}
\newcommand{\Wb}{\mathbf{W}}
\newcommand{\Xb}{\mathbf{X}}
\newcommand{\Yb}{\mathbf{Y}}
\newcommand{\Zb}{\mathbf{Z}}

\newcommand{\bA}{\bm{A}}
\newcommand{\bB}{\bm{B}}
\newcommand{\bC}{\bm{C}}
\newcommand{\bD}{\bm{D}}
\newcommand{\bE}{\bm{E}}
\newcommand{\bF}{\bm{F}}
\newcommand{\bG}{\bm{G}}
\newcommand{\bH}{\bm{H}}
\newcommand{\bI}{\bm{I}}
\newcommand{\bJ}{\bm{J}}
\newcommand{\bK}{\bm{K}}
\newcommand{\bL}{\bm{L}}
\newcommand{\bM}{\bm{M}}
\newcommand{\bN}{\bm{N}}
\newcommand{\bO}{\bm{O}}
\newcommand{\bP}{\bm{P}}
\newcommand{\bQ}{\bm{Q}}
\newcommand{\bR}{\bm{R}}
\newcommand{\bS}{\bm{S}}
\newcommand{\bT}{\bm{T}}
\newcommand{\bU}{\bm{U}}
\newcommand{\bV}{\bm{V}}
\newcommand{\bW}{\bm{W}}
\newcommand{\bX}{\bm{X}}
\newcommand{\bY}{\bm{Y}}
\newcommand{\bZ}{\bm{Z}}

\newcommand{\cA}{\mathcal{A}}
\newcommand{\cB}{\mathcal{B}}
\newcommand{\cC}{\mathcal{C}}
\newcommand{\cD}{\mathcal{D}}
\newcommand{\cE}{\mathcal{E}}
\newcommand{\cF}{\mathcal{F}}
\newcommand{\cG}{\mathcal{G}}
\newcommand{\cH}{\mathcal{H}}
\newcommand{\cI}{\mathcal{I}}
\newcommand{\cJ}{\mathcal{J}}
\newcommand{\cK}{\mathcal{K}}
\newcommand{\cL}{\mathcal{L}}
\newcommand{\cM}{\mathcal{M}}
\newcommand{\cN}{\mathcal{N}}
\newcommand{\cO}{\mathcal{O}}
\newcommand{\cP}{\mathcal{P}}
\newcommand{\cQ}{\mathcal{Q}}
\newcommand{\cR}{\mathcal{R}}
\newcommand{\cS}{{\mathcal{S}}}
\newcommand{\cT}{{\mathcal{T}}}
\newcommand{\cU}{\mathcal{U}}
\newcommand{\cV}{\mathcal{V}}
\newcommand{\cW}{\mathcal{W}}
\newcommand{\cX}{\mathcal{X}}
\newcommand{\cY}{\mathcal{Y}}
\newcommand{\cZ}{\mathcal{Z}}

\newcommand{\bell} {\bm{\ell}}
\newcommand{\balpha}{\bm{\alpha}}
\newcommand{\bbeta}{\bm{\beta}}
\newcommand{\bgamma}{\bm{\gamma}}
\newcommand{\bepsilon}{\bm{\epsilon}}
\newcommand{\bvarepsilon}{\bm{\varepsilon}}
\newcommand{\bzeta}{\bm{\zeta}}
\newcommand{\bvartheta}{\bm{\vartheta}}
\newcommand{\bkappa}{\bm{\kappa}}
\newcommand{\blambda}{\bm{\lambda}}
\newcommand{\bnu}{\bm{\nu}}
\newcommand{\bxi}{\bm{\xi}}
\newcommand{\bpi}{\bm{\pi}}
\newcommand{\bvarpi}{\bm{\varpi}}
\newcommand{\brho}{\bm{\varrho}}
\newcommand{\bsigma}{\bm{\sigma}}
\newcommand{\bvarsigma}{\bm{\varsigma}}
\newcommand{\btau}{\bm{\tau}}
\newcommand{\bupsilon}{\bm{\upsilon}}
\newcommand{\bphi}{\bm{\phi}}
\newcommand{\bvarphi}{\bm{\varphi}}
\newcommand{\bchi}{\bm{\chi}}
\newcommand{\bpsi}{\bm{\psi}}
\newcommand{\bomega}{\bm{\omega}}

\newcommand{\bGamma}{\bm{\Gamma}}
\newcommand{\bDelta}{\bm{\Delta}}
\newcommand{\bTheta}{\bm{\Theta}}
\newcommand{\bLambda}{\bm{\Lambda}}
\newcommand{\bXi}{\bm{\Xi}}
\newcommand{\bPi}{\bm{\Pi}}
\newcommand{\bSigma}{\bm{\Sigma}}
\newcommand{\bUpsilon}{\bm{\Upsilon}}
\newcommand{\bPhi}{\bm{\Phi}}
\newcommand{\bPsi}{\bm{\Psi}}
\newcommand{\bOmega}{\bm{\Omega}}

\newcommand{\RR}{\mathbb{R}}
\newcommand{\CC}{\mathbb{C}}

\def\R{\mathbb{R}}
\def\cov{\mathrm{cov}}
\def\pcov{\mathrm{Pcov}}
\def\pc{\mathrm{PC}}
\def\acos{\mathrm{arccos}}
\def \T{\mathrm{\scriptstyle T}} 
\def\sn {\sum_{i=1}^n}
\def \wt {\widetilde}
\newcommand {\summ}{\textnormal {sum}}
\newcommand{\argmin}{\mathop{\mathrm{argmin}}}
\newcommand{\argmax}{\mathop{\mathrm{argmax}}}

\newtheorem{theorem}{Theorem}[section]
\newtheorem{lemma}[theorem]{Lemma}
\newtheorem{remarka}[theorem]{Remark}
\newenvironment{remark}{\begin{remarka} \rm}{\end{remarka}}
\newtheorem{definition}[theorem]{Definition}
\newtheorem{assumption}{Assumption}
\newtheorem{corollary}[theorem]{Corollary}
\newtheorem{Conjecture}[theorem]{Conjecture}
\newtheorem{condition}[theorem]{Condition}

\begin{center}
    \bf{Abstract}
\end{center}
With the rapid development of quantum computers, quantum algorithms have been studied extensively. However, quantum algorithms tackling statistical problems are still lacking. In this paper, we propose a novel non-oracular quantum adaptive search (QAS) method for the best subset selection problems. QAS performs almost identically to the naive best subset selection method but reduces its computational complexity from $O(D)$ to $O(\sqrt{D}\log_2D)$, where $D=2^p$ is the total number of subsets over $p$ covariates. Unlike existing quantum search algorithms, QAS does not require the oracle information of the true solution state and hence is applicable to various statistical learning problems with random observations. Theoretically, we prove QAS attains any arbitrary success probability $q \in (0.5, 1)$ within $O(\log_2D)$ iterations. When the underlying regression model is linear, we propose a quantum linear prediction method that is faster than its classical counterpart. We further introduce a hybrid quantum-classical strategy to avoid the capacity bottleneck of existing quantum computing systems and boost the success probability of QAS by majority voting. The effectiveness of this strategy is justified by both theoretical analysis and extensive empirical experiments on quantum and classical computers.


\newpage


\section{Introduction}

The rapid advance in science and technology in the
past decade makes the quantum computer a reality,
offering researchers an unprecedented opportunity to
tackle more complex research challenges. 
Unlike classical computers depend on a classical bit having a {\it{state}} of either 0 or 1, a quantum computer operates on quantum processing units, quantum bits (or qubits), which can be in a state 0, 1, or both simultaneously due to the superposition property. The $n$ qubits create $2^n$ different states for the system that are superposed on each other. The superposition enables researchers to perform computations using all of those $2^n$ states simultaneously. In contrast, in classical computing, one conducts computations with one state at-a-time among these $2^n$ states. Consequently, one has to repeat the computations $2^n$ times in classical computing to get the quantum computation results.
\citep{nielsen2010quantum}.
Such a paradigm change has motivated significant developments of quantum algorithms in many areas, e.g.,  algebraic number theory \citep{shor1999polynomial, hallgren2002polynomial, van2008classical}, scientific simulations \citep{abrams1997simulation, byrnes2006simulating, kassal2008polynomial, wang2011quantum, hu2020quantum}, optimization \citep{jordan2005fast, jansen2007bounds, harrow2009quantum}, and many more.

The opportunity, however, has not yet been fully utilized by statisticians, because
there are significant challenges to developing quantum search algorithms to solve statistical problems.
First, a quantum computer does not provide deterministic results but instead gives a different result each time we run an algorithm according to some probability distribution.  For example, if we search the smallest number in a set of numbers using a classical computer, the result is determined if the algorithm is deterministic. However,  a quantum computer output the smallest number with a certain probability. That is, a quantum computer may output another number other than the correct result.
Second, existing quantum  search algorithms, in general,  depend on an oracle
function to decide if an outcome is a solution or not. For example, the celebrated Grover's algorithm \citep{grover1997quantum} requires an oracle function that can map all solutions to state 1 and all non-solutions to state 0. However,  such an oracle function is usually not available in practice. When the oracle is inaccurate, Grover's algorithm can perform as bad as a random guess which will be demonstrated in our numerical experiments.
Third, the current public available general-purpose quantum computer, such as the quantum computing system developed by IBM,  only has a few dozens of qubits, and thus allows very small piloting projects.

In this paper, we investigate the possibility of solving the best subset selection problem on a quantum computing system.  The best subset selection problem can be described as follows: Given a response $y$ and multiple predictors $x_1, x_2, \ldots, x_p$, we aim to find the best subset among $x_1, x_2, \ldots, x_p$ to fit the response $y$.  There are $2^p$ candidate subsets. For $p=30$, we have more than one billion candidate subsets.
Currently, there is no quantum algorithm specifically designed for the best subset selection problem. We propose a novel non-oracular quantum method named quantum adaptive search (QAS).
To overcome the aforementioned limitations, we randomly choose a model as our initial guess of the best subset.
QAS starts with an equally weighted superposition of the rest candidate models and iteratively updates the superposition towards the direction that minimizes the $\ell_2$ loss. Within each iteration, the new superposition is compared with the old one through a local evaluation function which does not require any oracle information of the true solution.  After the number of iterations, the result will replace our initial guess. We then repeat the above procedures until no more replacement can be made.
Since the output of the quantum computer is random,
we further introduce a hybrid quantum-classical strategy and boost the success probability of outputting the best subset by a majority voting.

Our theoretical and methodological contributions are as follows. First, we design an iterative algorithm to eliminate the requirement of an oracle function in the quantum algorithm. Second, we design a majority voting so that we have a high probability to get the correct solution from the algorithm.
Moreover, for the best subset selection problem over $D=2^p$ candidate models, within $O(\log_2D)$ iterations, QAS converges to a superposition that heavily weighs on the solution state and hence outputs the exact solution with a high probability.
In contrast,
a classic computing algorithm requires $O(D)$ quires to search for the best model. Last but not least,  the computational complexity of QAS is upper bounded by the order $O(\sqrt{D}\log_2D)$ which is only a $\log_2D$ factor larger than the theoretical lower bound for any oracular quantum search algorithms \citep{bennett1997strengths}.

Our empirical contribution is that we explore the simulation and real data analysis in IBM Quantum Experience\footnote{\url{www.ibm.com/quantum-computing}}, which is a publicly available cloud-based quantum computing system. The code and tutorial are publicly available. The exploration will facilitate the subsequent development in quantum algorithm development for solving statistical problems.  

The rest of the paper is organized as follows. We review the state-of-the-art quantum search algorithm and its limitation in Section \ref{sec:review}. In Section \ref{sec:best}, we propose a novel quantum adaptive search (QAS) algorithm and discuss its intuition and theoretical properties. We then introduce a quantum linear prediction algorithm for computing the criterion.
In Section \ref{sec:implement} , we introduce a quantum-classical hybrid strategy to improve the stability of quantum best subset selection. In Sections \ref{sec:exp_quantum} and \ref{sec:sim_classical}, we evaluate the empirical performance of the proposed method via extensive experiments on quantum and classical computers.
Section \ref{sec:conclusion} concludes the paper with a few remarks. Due to the limitation of space, the proofs of main theoretical results, additional simulation results, and a real data application are relegated to a supplemental material.


\section{Review of quantum search algorithm}
\label{sec:review}

\subsection{Notations and definitions}

To facilitate the discussion in the paper, we review some essential notations and definitions used in quantum computing that may be alien to the audience of statistics. We refer to \cite{nielsen2010quantum} and \cite{schuld2018supervised} for more detailed tutorials of quantum computing.
We start by introducing the linear algebra notations used in quantum mechanics, which were invented by Paul Dirac. In Dirac's notation,  a column vector ${\bf a}$ is denoted by $\ket{a}$ which reads as ${\bf a}$ in a `ket'. Typical vector space of interest in quantum computing is a Hilbert space $\cH$ of dimension $D=2^p$ with a positive integer $p$. 
For a vector $\ket{a} \in \cH$, we denote its dual vector as $\bra{a}$ (reads as $\bf a$ in a `bra'), which is an element of $\cH^*$. Besides, $\cH$ and $\cH^*$ together naturally induce an inner product $\braket{a|b} = \braket{{\bf a}, {\bf b}}$, which is also known as a `bra-ket'. We say $\ket{a}$ is a unit vector if $\braket{a|a}=1$. 


The framework of quantum computing resides in a state-space postulate which describes the state of a system by a unit vector in a Hilbert space. 
A state of a qubit can be described by a unit vector $\ket{\psi}$ in a two-dimensional Hilbert space. A superposition, as the linear combination of the states in one qubit, can be defined as
$$
\ket{\psi}=\phi_0\ket{0} + \phi_1 \ket{1},
$$
where coefficients $\phi_0$ and $\phi_1$ satisfy $|\phi_0|^2+ |\phi_1|^2=1$. Now suppose that we have two qubits and the system has four orthonormal states  $\ket{00}, \ket{01}, \ket{10}, \ket{11}$. Then, a superposition can be represented as 
$$
\ket{\psi}=\phi_0\ket{00} + \phi_1 \ket{01} + \phi_2\ket{10} + \phi_3 \ket{11},
$$
where coefficients $\phi_0$, $\phi_1$, $\phi_2$, $\phi_3$ satisfiy $|\phi_0|^2+ |\phi_1|^2+ |\phi_2|^2+ |\phi_3|^2=1$.
Analogously, a quantum computer of $p$ qubits can represent a state of a system by a unit vector $\ket{\psi}$ in a $D=2^p$ dimensional Hilbert space $\cH$. 
Let ${\cB} = \{\ket{b_i}\}_{i=0}^{D-1}$ be an orthonormal basis of $\cH$. Every superposition $\ket{\psi} \in \cH$ can be represented as
\begin{equation}
   \ket{\psi}  = \sum\limits_{i=0}^{D-1} \phi_i \ket{b_i},  
   \label{eq:psi}
\end{equation}
where $\phi_0 ,\ \ldots, \  \phi_{D-1}$ is a set of coefficients with $\phi_i = \braket{b_i|\psi}$ and $\sum_{i=0}^{D-1} |\phi_i|^2=1$.


Another salient feature of quantum computing is the measurement of a quantum system yields a probabilistic outcome rather than a deterministic one. Suppose the superposition $\ket{\psi}$ in \eqref{eq:psi} is measured, it collapses to a random state in ${\cB}$. In addition, the probability we observe $\ket{b_i}$ is $|\phi_i|^2$ for $i=0,\ldots, D-1$.

Different from a classical computer with logic gates, e.g., AND and NOT gates, a quantum computer is built with quantum gates. 
These quantum gates not only have the logic gates mentioned above but also include the new gates that are not available in classical computers.
For example, the Hadamard gate enables the rotations and reflections of the input states. Some quantum algorithms that make use of these new gates have shown some remarkable results that significantly outperform the algorithms in classical computers.

\subsection{Grover's algorithm}\label{sec:grover}

The best subset selection problem can be decomposed into two subproblems: (1) What is the criterion for the best? (2) Given the criterion, which subset is the best?  For the first subproblem, there are many criteria, e.g, AIC, BIC, cross-validation, and mean squared error. For the second subproblem, searching for the best is challenging if the number of candidate subsets is super-large.  We shall review Grover's algorithm, which is the most popular quantum searching algorithm for the second subproblem.

Recall that given $p$ predictors, there are  $D=2^p$ candidate subsets, from which we aim to find the best subset.  Once the best criterion is chosen and applied to each subset, we have $D=2^p$  criterion values computed for all subsets. We aim to find the solution (e.g. the smallest or the largest value) from $D=2^p$ elements, which are indexed by $0, 1, \ldots, D-1$. To ease the presentation, we assume there are no tied solutions although the discussions in this paper can be easily generalized to such cases.
In the language of quantum computing, each index is regarded as a state, which can be modeled by a unit vector in an orthonormal basis ${\cal D}=\{\ket{i}\})_{i=0}^{D-1}$ of a $D$ dimensional Hilbert space $\cH$.


\subsubsection{ Geometric view of Grover's algorithm}
\begin{wrapfigure}[10]{r}{0.50\textwidth} 
  \vspace{-30pt}
  \begin{center}
    \includegraphics[width=0.45\textwidth]{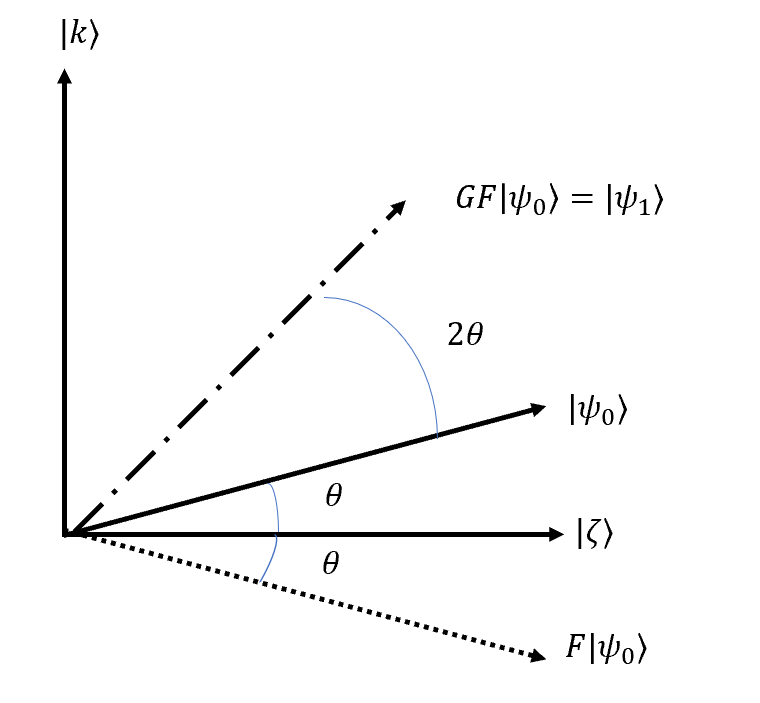}
    \caption{Geometrical visualization of the two steps in the first Grover's operation. }
    \label{fig:geo_view}
  \end{center}
\end{wrapfigure}

We now provide an intuitive geometric presentation of Grover's algorithm which is summarized in Algorithm \ref{algo:grover} below. Denote the solution state by $\ket{k}$ and the average of all non-solution states by $\ket{\zeta}= \frac{1}{\sqrt{D-1}}\sum\limits_{i\neq k} \ket{i}$.
Notice that $\ket{\zeta}$ is orthogonal to the solution state $\ket{k}$. We plot them as $x$-axis and $y$-axis in Figure~\ref{fig:geo_view}.
Grover's algorithm is initialized with a superposition as the equally weighted average of all states $\ket{i}$, $i=0, \ldots , D-1$. To be specific, the initial superposition is defined as
\begin{align*}
    \ket{\psi_0}=\frac{1}{\sqrt{D}} \sum\limits_{i=0}^{D-1} \ket{i}. 
\end{align*}
Denote the angel between the initial superposition $\ket{\psi_0}$ and $\ket{\zeta}$ by $\theta$. It is easy to see that $\sin\theta =\frac{1}{\sqrt{D}}$.

The first iteration of Grover's algorithm contains two operations. The first operation 
reflects $\ket{\psi_0}$ with respect to $\ket{\zeta}$ to get  $F\ket{\psi_0}$,
The second operation
 reflects $F\ket{\psi_0}$ with respect to $\ket{\psi_0}$ to get $GF\ket{\psi_0}$, which is our new superposition, denoted by $\ket{\psi_1}$. As a result, the angle between $\ket{\psi_1}$ and $\ket{\zeta}$ is $3\theta$. 
Figure \ref{fig:geo_view} provides a visualization of the two operations in the first Grover's operation. Then, Grover's algorithm iteratively repeats the above two operations, i.e, a reflection w.r.t.  $\ket{\zeta}$  and then a reflection w.r.t. $\ket{\psi_0}$. After the $j$th iteration, the the angle between $\ket{\psi_j}$ and $\ket{\zeta}$ is $(2 j + 1)\theta$. The iteration terminates if the $j$th state $\ket{\psi_j}$ coincides with the solution state  $\ket{k}$, i.e.,  $\ket{\psi_j}=\ket{k}$, which implies that  
\begin{equation}\label{jthgrover}
    (2 j + 1)\theta = \pi/2.
\end{equation}


The typical motivation to use quantum search algorithm is that the number of candidate set $D$ is large.
When $D$ is large, i.e., $\theta$ is small, we can approximate the angle by 
\begin{equation}\label{approtheta}
    \theta \approx \sin\theta =\frac{1}{\sqrt{D}}.
\end{equation}
Substituting (\ref{approtheta}) into (\ref{jthgrover}), we have $(2 j+1)/\sqrt{D}=\pi/2$, which yields $j$ is approximately $ \sqrt{D}\pi/4$.  
This is a remarkable result since the number of iteration is in the order of $O(\sqrt{D})$.
Grover's algorithm can be easily extended to a search problem with $M > 1$ solutions, where the number of iteration $j$ is approximately $ \sqrt{D/M}\pi/4 $. See \cite{boyer1998tight} for more detailed discussions.



\subsubsection{ Technical details of Grover's algorithm}
Now we present Grover's algorithm with more mathematical rigor. 
Grover's algorithm assumes that we are provided with an quantum oracle evaluation function $S(\cdot)$, which can recognize the solution to the search problem. That is, suppose that index $k \in \{0, \ldots , D-1\}$ corresponds to the best subset, we have  $S(\ket{k})=1$ and $S(\ket{i})=0$ for $i\neq k$.
Recall that, Grover's algorithm is initialized with the following superposition
\begin{align*}
    \ket{\psi_0}=\frac{1}{\sqrt{D}} \sum\limits_{i=0}^{D-1} \ket{i} \equiv c_0\ket{k} + d_0 \sum\limits_{i\neq k} \ket{i},
\end{align*}
where $c_0 = d_0 = \frac{1}{\sqrt{D}}$. Then, Grover's algorithm updates $c$ and $d$ in an iterative manner. In the $j$th iteration,  Grover's algorithm applies the following two operations to the current superposition $\ket{\psi_{j-1}} = c_{j-1}\ket{k} + d_{j-1} \sum\limits_{i\neq k} \ket{i}$,

\begin{itemize}
    \item[(a)] A flip operation $F$ to $\ket{\psi_{j-1}}$, where $F\ket{i}=\left(1-2S(\ket{i})\right)\ket{i}$. In other words,   $F\ket{k}=-\ket{k}$ and $F\ket{i}=\ket{i}$ for $i \neq k$.
    \item[(b)] A Grover's diffusion operation $G$ to $F\ket{\psi_{j-1}}$, where $G = 2\ket{\psi_{0}}\bra{\psi_{0}}-{\bI}_D$ and ${\bI}_D$ is a $D\times D$ identity matrix.
\end{itemize}

In the rest of this paper, we call the two operations together, i.e. $GF$, as Grover's operation

After the $j$th iteration, the superposition $\ket{\psi_{j-1}}$ is updated to 
$$\ket{\psi_{j}}=GF\ket{\psi_{j-1}}= c_{j}\ket{k} +  d_{j}\sum\limits_{i\neq k} \ket{i},$$
where the coefficients $c_j$ and $d_j$ satisfy
\begin{align*}
    \begin{cases} c_{j} = \frac{D-2}{D}c_{j-1} + \frac{2(D-1)}{D}d_{j-1}, \\
                  d_{j} =  \frac{D-2}{D}d_{j-1} - \frac{2}{D}c_{j-1}.
    \end{cases}
\end{align*}

Let $\theta$ be the angle that satisfies $\sin^2\theta = \frac{1}{D}$. The coefficients $c_j$ and $d_j$ admit a closed form \citep{nielsen2010quantum},
\begin{align*}
    \begin{cases} c_{j} = \sin\big((2j+1)\theta\big), \\
                  d_{j} =  \frac{1}{\sqrt{D-1}}\cos\big((2j+1)\theta\big).
    \end{cases}
\end{align*}

\begin{algorithm}
    \caption{Grover's algorithm}
    \label{algo:grover}
    \begin{algorithmic}
        \State \textbf{Input}: An orthonormal basis $\mathcal{D}$ of size $D=2^p$, a binary evaluation function $S$ associated with an oracle state $|k\rangle$, such that $S(\ket{k})=1$ and $S(\ket{i})=0$ for $i\neq k$.
        Number of iterations $\tau=\lceil \pi \sqrt{D} /4\rceil$, where $\lceil \cdot \rceil$ is the ceiling function.
        \State \textbf{Initialization} Prepare an equally weighted superposition on a quantum register of $p$-qubits  $\ket{\psi_0} =  \frac{1}{\sqrt{D}}\sum_{i=0}^{D-1} \sti$.
        \For{$j=1,\ldots, \tau$}
        \State  Grover's operation: Let $\ket{\psi_{j}}=GF\ket{\psi_{j-1}}$, where $F\ket{i}=\left(1-2S(\ket{i}\right)\ket{i}$,  
        \State $G = 2\ket{\psi_{0}}\bra{\psi_{0}}-{\bI}_D$, and ${\bI}_D$ is a $D\times D$ identity matrix.
        \EndFor
        \State \textbf{Output}: Measure the latest superposition $\ket{\psi_{\tau}}$ on the quantum register.
    \end{algorithmic}
\end{algorithm}

\subsubsection{Significance and challenges of Grover's algorithm}
\label{sec:limitation}

\cite{grover1997quantum} proposed a quantum search algorithm with a success probability of at least 50\%. The computational complexity of Grover's algorithm is of the order $O(\sqrt{D})$.  \cite{bennett1997strengths}  showed that this rate is optimal, up to a constant, among all possible quantum search algorithms. In contrast, any classical search algorithm needs to query the system for at least $0.5D$ times to solve the searching problem with a 50\% or higher success probability. As a result, most classical search algorithms have the computational complexity of the order $O(D)$.


Grover's algorithm is a quantum algorithm that depends on an oracle evaluation function that maps the solution state to 1 and all other states to 0. However, such a piece of oracle information is usually not available in statistical problems including the best subset selection.   When we have no oracle information of the solution state at all, the oracle evaluation function can only be constructed by a randomly selected solution state. Then, Grover's algorithm is highly likely to rotate the initial superposition in a wrong direction and the output of the algorithm can be as bad as a random guess. We empirically demonstrate this phenomenon in Section \ref{sec:exp_QAS}.



\section{Best subset selection with quantum adaptive search}\label{sec:best}

In this section, we first review the best subset selection problem. We then propose a novel quantum search algorithm named quantum adaptive search (QAS) which aims to overcome the aforementioned limitation of Grover's algorithm. We also provide a quantum linear prediction (QLP) method to compute the loss function.

\subsection{Review of best subset selection}
Best subset selection has been a classical and fundamental method for linear regressions \citep[e.g.][]{hocking1967selection, beale1967discarding, shen2012likelihood,fan2020best}. When the covariates contain redundant information for the response, selecting a parsimonious best subset of covariates may improve the model interpretability as well as the prediction performance. 
The best subset selection for linear regression can be formulated as
\begin{equation}
    \min_{\bbeta \in \RR^p} \| \bY - \bX\bbeta\|_2^2 \quad \text{subject to} \quad \|\bbeta\|_0\leq t,
    \label{eq:L0}
\end{equation}
where $\bY=(y_1, \ \ldots, \ y_n)^{\T} \in \RR^n$ is a response vector, $\bX = (x_{i,j})_{1\leq i\leq n, 1\leq j \leq p} \in \RR^{n\times p}$ is a design matrix of  covariates and $\bbeta = (\beta_1, \ \ldots, \  \beta_p)^{\T} \in \RR^p$ is a vector of unknown regression coefficients. Throughout this paper, we assume $\bY$ and $\bX$ are centered for the simplicity of presentation. 
Further, $\|\bbeta\|_0$ is the $\ell_0$ norm of $\bbeta$ and $0\leq t \leq \min{(n,p)}$ is a subset size.
In general, solving \eqref{eq:L0} is a non-convex optimization problem. Without a convex relaxation or a stage-wise approximation, the optimization involves a combinatorial search over all subsets of sizes $\{1, \ \ldots, \  \min(n,p)\}$ and hence is an NP-hard problem \citep{welch1982algorithmic, natarajan1995sparse}. Therefore, seeking the exact solution of \eqref{eq:L0} is usually computationally intractable when $p$ is moderate or large.

To overcome the computational bottleneck of the best subset selection, many alternative and approximate approaches have been developed and carefully studied. Stepwise regression methods \citep[e.g.][]{efroymson1960multiple, draper1966applied} sequentially add or eliminate covariates according to their marginal contributions to the response condition on the existing model. Stepwise regression methods speed up the best subset selection as they only optimize \eqref{eq:L0} over a subset of all candidates, and hence their solutions may not coincide with the solution of \eqref{eq:L0}. However, stepwise regression methods suffer from the instability issue in high dimensional regime \citep{breiman1996heuristics}. 
Regularized regression methods suggest to relax the nonconvex $\ell_0$ norm in \eqref{eq:L0} by various convex or nonconcave alternatives. A celebrated member in the house is LASSO \citep{tibshirani1996regression} which solves an $\ell_1$ regularized problem instead. Elegant statistical properties and promising numerical performance soon made regularized regression a popular research topic in statistics, machine learning and other data science related areas. We refer to basis pursuit \citep{chen1994basis}, SCAD \citep{fan2001variable}, elastic net \citep{zou2005regularization},  adaptive LASSO \citep{zou2006adaptive}, Dantzig selector \citep{candes2007dantzig}, nonnegative garrotte \citep{yuan2007non}, and MCP \citep{zhang2010nearly}, among many others. Though regularized regression methods can perform equivalent to or even better than the solution of \eqref{eq:L0} in certain scenarios \citep[e.g.][]{zhao2006model, fan2014strong, hastie2017extended}, they are not universal replacements for the best subset selection method. Nevertheless, solving  \eqref{eq:L0} efficiently remains a statistically attractive and computationally challenging problem.


Recently, \cite{bertsimas2016best} has reviewed the best subset selection problem from a modern optimization point of view.  Utilizing the algorithmic advances in mixed integer optimization,  \cite{bertsimas2016best} proposed a highly optimized solver of \eqref{eq:L0} which is scalable to relatively large $n$ and $p$.  Later, \cite{hazimeh2018fast} considered a new hierarchy of necessary optimality conditions and propose to solve \eqref{eq:L0} by adding extra convex regularizations. Then, \cite{hazimeh2018fast} developed an efficient algorithm based on coordinate descent and local combinatorial optimization. Such recent optimization developments push the frontier of computation for the best subset selection problems by a big margin.

\subsection{Quantum adaptive search algorithm}
\label{sec:qas}

Let us consider the best subset selection problem \eqref{eq:L0} and assume $n\geq p$ such that all $D=2^p$ subsets of $p$ covariates are attainable. When $n<p$, 
we only need to search over $\sum_{t=0}^n{p \choose t}<2^p$ subsets which is a less challenging combinatorial search problem.

\subsubsection{Overview of quantum adaptive search algorithm}
\label{sec:overqas}


Recall that each subset of $\{1, \ \ldots, \ p\}$ can be encoded as a state in an orthonormal basis $\mathcal{D} = \{\ket{0}, \ \ldots, \  \ket{D-1} \}$, which consists of $D = 2^p$ elements of a Hilbert space $\cH$. 
We define a state loss function $g(\cdot): \cH \rightarrow \R$ by the mean squared error (MSE), 
\begin{equation}
    g(\ket{i}) \equiv L_n(t; j_1, \ \ldots, \ j_t; \widehat{\beta}_{j_1}, \ \ldots \ ,  \widehat{\beta}_{j_t} ) = \frac{1}{n}\sum\limits_{i=1}^n \left(y_i - \sum_{l=1}^t\widehat\beta_{j_l} x_{i,j_l} \right)^2,
    \label{eq:evaluation}
\end{equation}
where the state $\ket{i} \in \mathcal{D}$ is a vector in $\cH$ that corresponds to the subset $\{j_1, \ \ldots, \ j_t\}$, and  $\widehat{\beta}_{j_1}, \ \ldots \ ,  \widehat{\beta}_{j_t}$ are the regression coefficient estimates obtained by minimizing \eqref{eq:L0} with fixed $t$ and $\{j_1, \ \ldots, \ j_t\}$. Intuitively, the best subset corresponds to the state that minimizes the state loss function.

Now, we present the proposed non-oracular quantum adaptive search method, i.e. QAS, by three key steps as follows.

\noindent{\sc{(1) Initialization}:} We choose an initial benchmark state $\ket{w}$ randomly from the set $\mathcal{D} = \{\ket{0}, \ \ldots, \  \ket{D-1} \}$. Also, we pre-specify a learning rate $\lambda \in (0,1)$.

\noindent{\sc{(2) Updating}:} We run Algorithm \ref{algo:grover} on set $\mathcal{D}$ by inputting the benchmark state $\ket{w}$ and the number of iterations $\tau=\lceil \pi \lambda^{-m/2} /4\rceil$, where $m$ is a positive integer. Denote the output of Algorithm \ref{algo:grover} as $\ket{w^{new}}$ which is a state in $\mathcal{D}$. Then, we compare $\ket{w^{new}}$ with $\ket{w}$ in terms of the state loss function $g(\cdot)$. If $g(\ket{w^{new}})<g(\ket{w})$, we update the current benchmark state $\ket{w}$ with  $\ket{w^{new}}$, otherwise we do not update $\ket{w}$.

\noindent{\sc{(3) Iteration and output}:} Start with $m=1$ and repeat the updating step. After each updating step, set $m=m+1$. The QAS stops when $m>C(\lambda) \ln{D}$, where $C(\lambda)$ is a positive constant that depends on the learning rate $\lambda$. Then, we measure the quantum register with the latest superposition. The output is the observed state in $\mathcal{D}$ and its corresponding subset.

Unlike Grover's algorithm, QAS  randomly selects a state in  $\mathcal{D}$ as the benchmark state which does not require any oracle information of the solution state. Then,  QAS iteratively updates the benchmark state towards the direction that reduces the state loss function. Besides, QAS is data-adaptive in the sense that it starts with a conservative learning step size (e.g. $m=1$) and gradually increases the learning step size as the benchmark state has been updated towards the truth. We summarize QAS in Algorithm \ref{algo:adaptive_search} below.

\begin{algorithm}
    \caption{Quantum adaptive search}
    \label{algo:qas}
    \begin{algorithmic}
        \State \textbf{Input}: An orthonormal basis $\mathcal{D}$ of size $D=2^p$, a state loss function $g(\cdot)$ that maps a state in $\mathcal{D}$ to a real number, a learning rate $\lambda \in (0, 1)$.
        \State \textbf{Initialization} Set $m=1$. Randomly select a state in $\mathcal{D}$ as the  initial benchmark state $\ket{w}$. Define a local evaluation function $S(\ \cdot \  , \ket{w}, g)$ such that $S(\ket{i}, \ket{w}, g)=1$ if $g(\ket{i})\leq g(\ket{w})$ and $S(\ket{i}, \ket{w}, g)=0$ if $g(\ket{i}) > g(\ket{w})$. 
        \Repeat
        \State (1) Run Algorithm \ref{algo:grover} by inputting $\cD$, $S(\ \cdot\ , \ket{w}, g)$ and $\tau(m)=\lceil \pi \lambda^{-m/2} /4\rceil$.
        \State (2) Measure the quantum register and denote the readout by $\ket{w^{new}}$.
        \State (3) If $g(|w^{new}\rangle)<g(|w\rangle)$, set $|w\rangle = |w^{new}\rangle$ and update $S(\ \cdot \ , \ket{w}, g)$ accordingly.
        \State (4) $m=m+1$.
        \Until{$ m > C(\lambda)\ln{D} $}, where $C(\lambda)$ is a positive constant depends on $\lambda$.
        \State \textbf{Output}: The latest benchmark state $\ket{w}$.
    \end{algorithmic}
    \label{algo:adaptive_search}
\end{algorithm}

Algorithm \ref{algo:adaptive_search} provides a general non-oracular quantum computing framework for best subset selection problems as the state loss function $g(\cdot)$ can be tailored for various statistical models, such as nonlinear regression, classification, clustering, and low-rank matrix recovery.

\subsubsection{Intuition of quantum adaptive search}\label{sec:qas_intuition}
In this subsection, we demonstrate the intuition of QAS.
Suppose that we implement QAS on an orthonormal basis $\mathcal{D} = \{\ket{i}\}_{i=0}^{D-1}$ with a pre-specified state loss function $g(\cdot): \cH \rightarrow \R$ and a learning rate $\lambda \in (0,1)$.  We assume there exists a sole solution state in $\cD$ that minimizes $g(\cdot)$. 
Without loss of generality, we number the states in $\cD$ as the ascending rank of their state loss function values, i.e.
\begin{equation}
    g(\ket{0}) <   g(\ket{1}) \leq g(\ket{2}) \leq \ \cdots \ \leq g(\ket{D-1}),
    \label{eq:g_rank}
\end{equation}
where $\ket{0}$ is the sole solution state.

If the initialization step in Algorithm \ref{algo:qas} luckily selects the sole solution state $\ket{0}$ as the initial benchmark state, QAS will never update the benchmark state and hence reduces to Grover's algorithm with a known oracle state and $\tau \approx \sqrt{D} \pi  /4 $. Therefore, QAS can recover the true state with a high success probability. 

A more interesting discussion would be considering the initial benchmark state $\ket{w}$ does not coincide with the truth, i.e. $\ket{w} \neq \ket{0}$. 
Given the rank in \eqref{eq:g_rank}, the local evaluation function $S(\ket{i}, \ket{w}, g)$ can be simplified as
\begin{align*}
    \begin{cases} S(\ket{i} , \ket{w}, g) =   1,  \quad \text{ if } i\leq w,\\
                  S(\ket{i} , \ket{w}, g) =   0,  \quad \text{ if } i>w.
    \end{cases}
\end{align*}

In the $m$th iteration, QAS calls Algorithm \ref{algo:grover} by inputting  $\cD$, $S(\ \cdot\ ,\ket{w}, g)$ (suppose that $w\neq 0$) and $\tau(m)=\lceil \pi \lambda^{-m/2} /4\rceil$. Algorithm \ref{algo:grover} initializes a equally weighted superposition as
\begin{align*}
    \ket{\psi_0}=\frac{1}{\sqrt{D}} \sum\limits_{i=0}^{D-1} \ket{i} \equiv \alpha_0\sum\limits_{i=0}^w \ket{i}+  \beta_0 \sum\limits_{j=w+1}^{D-1} \ket{j}, \quad \text{with} \quad \alpha_0=\beta_0=\frac{1}{\sqrt{D}}.
\end{align*}
Then, Algorithm \ref{algo:grover} applies $\tau(m)$ Grover's operations to $\ket{\psi_0}$ which updates $\ket{\psi_{0}}$ to $\ket{\psi_{\tau(m)}}$ with coefficients satisfy
\begin{align*}
    \begin{cases} \alpha_{\tau(m)} = \frac{1}{\sqrt{w+1}}\sin\big((2\tau(m)+1)\theta\big), \\
                  \beta_{\tau(m)} =  \frac{1}{\sqrt{D-w-1}}\cos\big((2\tau(m)+1)\theta\big),
    \end{cases}
\end{align*}
where the angle $\theta$ satisfies $\sin^2\theta = (w+1)/D$. After the $m$th iteration, Algorithm \ref{algo:grover} outputs a random state $\ket{w_{new}} \in \cD$ with the following probability mass function
\begin{align*}
    P(\ket{w_{new}}=\ket{i}) = \begin{cases} \alpha_{\tau(m)}^2,  \quad \text{ if } i \leq w,\\
                  \beta_{\tau(m)}^2,  \quad \text{ if } i>w.
    \end{cases}
\end{align*}

\begin{figure}[ht]
    \centering
    \includegraphics[width=\textwidth]{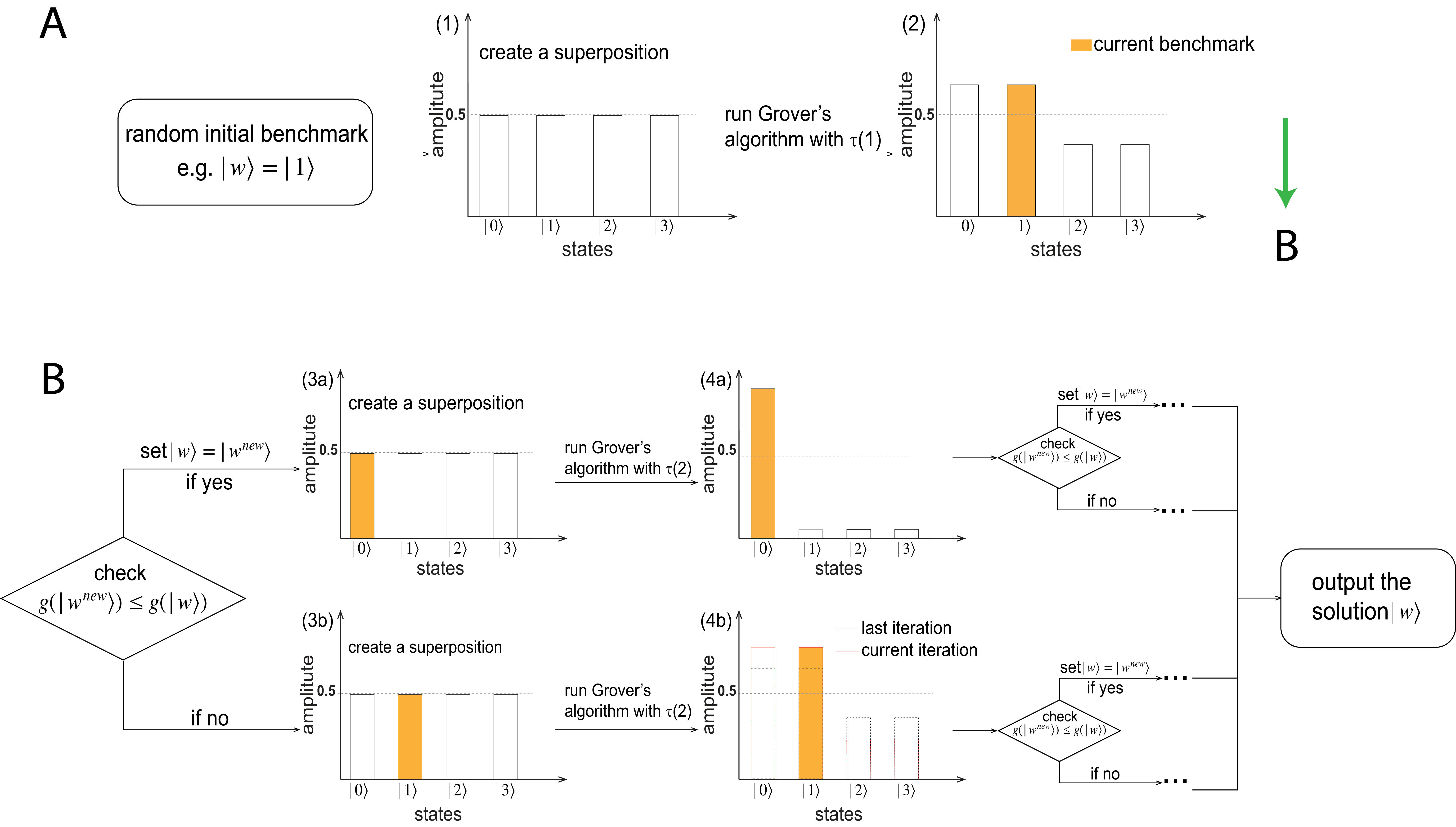}
    \caption{Illustrative example of quantum adaptive search.}
    \label{fig:FlowChart1}
\end{figure}

Since the  learning rate $\lambda\in (0,1)$, we have $\alpha_{\tau(m)}^2 > 1/D > \beta_{\tau(m)}^2$ for some positive $m$. After the $m$th iteration, QAS amplifies the probability of drawing the states whose state loss function values are smaller or equal to $g(\ket{w})$. Meanwhile, QAS suppresses the probability of drawing the states whose state loss function values are greater than $g(\ket{w})$. Geometrically, QAS rotates the initial superposition $\ket{\psi_0}$ towards  $\frac{1}{\sqrt{w+1}} \sum\limits_{i=0}^w \ket{i}$, which is an average over $\ket{w}$ and the states that can reduce the state loss function from  $\ket{w}$. If the output of the $m$th iteration is $\ket{w_{new}}=\ket{i}$ for some $i\geq w$, QAS will not update $\ket{w}$. On the other hand, if  $i< w$, QAS will update $\ket{w}$ with $\ket{w_{new}}$ which is equivalent to descend $\ket{w}$ to a state with a smaller state loss function value. In Figure \ref{fig:FlowChart1}, we visually illustrate the mechanism of QAS when $p=2$.

\subsection{Theoretical analysis of quantum adaptive search}\label{sec:qas_theory}

Intuitively, one can think of QAS as a ``quantum elevator'' that starts at a random floor of a high tower and aims to descent to the ground floor. In each operation, a quantum machine will randomly decide if this elevator stays at the current floor or goes down to a random lower-level floor. Besides, the probability of going down will gradually increase as the number of operations increases. Thus, it is not hard to imagine that the ``quantum elevator'' can descent to the ground floor with a high success probability after a large enough amount of operations. Our intuition of QAS is justified by the following theorem. Theorem \ref{thm:QAS_main} states that within $O(\log_2 D)$ iterations, QAS finds the sole solution state for any pre-specified success probability greater than 50\%. 

\begin{theorem}\label{thm:QAS_main}
Let $\kappa \in (0.5, 1)$ be a constant. With probability at least $\kappa$, the quantum adaptive search (e.g. Algorithm \ref{algo:qas}) finds the sole solution state $\ket{0}$ within $C_{\kappa}\log_2{D}$ iterations, where $C_{\kappa}$ is a positive constant that depends on $\kappa$.
\end{theorem}

Suppose that, in the $m$th iteration of Algorithm \ref{algo:qas}, the current benchmark state is $\ket{w_{m}}$ with $g(\ket{w_m}$ being the $r_m$th smallest among $\{g(\ket{i})\}_{i=0}^{D-1}$. Theorem \ref{thm:QAS_iter} below shows the expected number of Grover's operations that Algorithm \ref{algo:qas} needs to update $g(\ket{w_m})$ is on the order $O(\sqrt{D/r_m})$. This together with Theorem \ref{thm:QAS_main} imply the computational complexity upper bound of QAS is of order $O(\sqrt{D}\log_2D)$ which is only a $\log_2D$ factor larger than the theoretical lower bound of any oracular quantum search algorithm \citep{bennett1997strengths}. Notice that, QAS achieves a near optimal computational efficiency as oracular quantum search algorithms without using any oracle information.

\begin{theorem}\label{thm:QAS_iter}
Let $\ket{w_{m}}$ be the current benchmark state in the $m$th iteration of Algorithm \ref{algo:qas}. Let $r_m$ be the rank of $g(\ket{w_{m}})$ in the sorted  sequence of $\{g(\ket{i})\}_{i=0}^{D-1}$ in ascending order, $r_m \in \{1,\ldots ,D\}$. The expected time for Algorithm \ref{algo:qas} to update $\ket{w_{m}}$ is of order $O(\sqrt{D/r_m})$.
\end{theorem}

\subsection{Computation of the loss function via quantum computing}
\label{sec:qlp}

The success of QAS relies on the efficient comparison of two states through the state loss function defined in  \eqref{eq:evaluation}, which is the MSE of linear regression.
Recently, a line of studies \citep{wiebe2012quantum, rebentrost2014quantum, wang2017quantum, zhao2019quantum} focuses on formulating linear regression as a matrix inversion problem, which can be tackled by the quantum algorithm to solve linear systems of equations \citep{harrow2009quantum}. Their goal is to find a series of quantum states whose amplitudes represent the ordinary least squares estimator of linear regression coefficients. Although their algorithms can encode such quantum states efficiently, ``it may be exponentially expensive to learn via tomography” \citep{harrow2009quantum}. As a result, most existing quantum linear regression algorithms require the input design matrix to be sparse. Otherwise, the computation is slow and the estimation accuracy may suffer from noise accumulation.

In this 
paper, we investigate a novel quantum linear prediction approach that is based on the singular-value decomposition. The proposed method avoids the matrix inversion problem. Instead, we estimate the inverse singular values by utilizing a recently developed quantum tomography technique \citep{lloyd2014quantum}.
Let $\{j_1, \ \ldots, \ j_d\}$ be a subset of $\{1, \ \ldots, \ p\}$ with $d\leq n$. Denote $\bX_d \in \RR^{n\times d}$ the sub-matrix of $\bX$ corresponding to the subset $\{j_1, \ \ldots, \ j_d\}$. Let $\widetilde{\bx}$ be a new  observation of the $d$ selected covariates. Then, the corresponding response value $\widetilde{y}$ can be predicted by 
\begin{equation}
    \widehat{y} = \widetilde{\bx}^{\T} (\bX_d^{\T}\bX_d)^{-1}\bX_d^{\T}\by.
    \label{eq:yhat}
\end{equation}

Suppose that the rank of $\bX_d$ is $R \leq d$, a compact singular value decomposition of $\bX_d$ yields
\begin{equation*}
    \bX_d = \bU \bSigma \bV^{\T}=\sum\limits_{r=1}^R \sigma_r \bu_r \bv_r^{\T},
    \label{eq:svd}
\end{equation*}
where $\bSigma=\mathrm{diag}\{\sigma_1, \ \ldots, \ \sigma_R\}$ is a diagonal matrix of $R$ positive singular values, and $\bU=(\bu_1, \ \ldots, \  \bu_R)\in\RR^{n\times R}$ and $\bV=(\bv_1, \ \ldots, \  \bv_R)\in\RR^{d\times R}$ are matrices of left and right singular vectors, respectively. Further, we have $\bU\bU^{\T}=\bV\bV^{\T}=\bI_R$. Then, the linear predictor in \eqref{eq:yhat} can be represented by
\begin{equation}
    \widehat{y} = \sum\limits_{r=1}^R \sigma_r^{-1} \widetilde{\bx}^{\T} \bv_r \bu_r^{\T} \by.
    \label{eq:yhat_svd}
\end{equation}

Motivated by the quantum principle component analysis (QPCA) \citep{lloyd2014quantum}
and the inverted singular value problem \citep{schuld2016prediction}, we propose to calculate \eqref{eq:yhat_svd} by a quantum linear prediction (QLP) procedure, which can be summarized as the following three steps.

\noindent{\sc{Step 1: Quantum states preparation.}}

We encode $\bX_d$, $\bv_r$ and $\bu_r$, $r=1, \ \ldots, \  R$, into  a quantum system as
\begin{equation}
    \ket{\bX_d} = \sum_{j=1}^{d} \sum_{i=1}^{n} a_{i,j} \ket{i} \ket{j},
     \quad
    \ket{\bu_r} = \sum\limits_{i=1}^{n}  \mu_{r,i}\ket{i} \quad \text{and} \quad
    \ket{\bv_r} = \sum\limits_{j=1}^{d}  \nu_{r,j}\ket{j},
    \label{eq:psi_X}
\end{equation}
where $\sum_{i}\sum_j|a_{i,j}|^{2}=\sum_{i}| \mu_{r,i}|^{2}=\sum_{j}|\nu_{r,j} |^{2}=1$, and
$\{\ket{i}\}_{i=1}^n$ and $\{\ket{j}\}_{j=1}^d$ are orthonormal quantum bases representing the linear space spanned by the left and right singular vectors of $\bX_d$, respectively.

Then, we encode $\bX_d^{\T}\bX_d$ and the compact singular value decomposition of $\bX_d$ into  a quantum system as
\begin{equation}
    {\mathbf \rho} \equiv {\mathbf \rho}(\bX_d^{\T}\bX_d) = \sum_{j, j'=1}^{d} \sum_{i=1}^{n} a_{i,j} a_{i,j'}\ket{j} \bra{j'} 
    \quad \text{and} \quad
    \ket{\bX_d}=\sum\limits_{r=1}^{R} \sigma_{r}\ket{\bu_r}\ket{\bv_r}.
    \label{eq:rho}
\end{equation}

Similarly, we can encode $\by$ and $\widetilde{\bx}$ into  a quantum system as
\begin{align}
   \left|\mathbf{y}\right\rangle = \sum_{\eta=1}^{n} b_{\eta} \ket{\eta} \quad \text{and} \quad
    \ket{\widetilde{\bx}} = \sum_{\gamma=1}^{d} c_{\gamma} \ket{\gamma},
    \label{eq:y_x}
\end{align}
with $\sum_{\eta}|b_{\eta}|^{2}=\sum_{\gamma}|c_{\gamma} |^{2}=1$, and $\{\ket{\eta}\}_{\eta=1}^n$ and $\{\ket{\gamma}\}_{\gamma=1}^d$ are orthonormal quantum bases.

\vspace*{0.1in}

\noindent{\sc{Step 2: Extracting the inverted singular values.}}

To calculate \eqref{eq:yhat_svd} with a quantum computer, we first need to calculate the inverse of singular values of $\bX_d$. According to the ideas of QPCA  \citep{lloyd2014quantum}, we apply ${\mathbf \rho}$ to $\ket{{\bX_d}}$ and follow the quantum phase estimation algorithm \citep[e.g.][]{shor1994algorithms, szegedy2004quantum, wocjan2009quantum}  which yields
\begin{equation}
    \sum_{r=1}^{R} \sigma_{r} \ket{\bv_r} \ket{\bu_r}\ket{\lambda_r},
    \label{eq:sigma_r}
\end{equation}
where $\lambda_{r}=\sigma_{r}^{2}$, $r=1, \ \ldots, \  R$, are the eigenvalues of ${\mathbf \rho}$ which is encoded in a quantum register $\ket{\lambda_r}$.  Then, by adding an extra qubit and rotating \eqref{eq:sigma_r} condition on the eigenvalue register $\ket{\lambda_r}$, we have
\begin{equation}
    \sum_{r=1}^{R} \sigma_{r} \ket{\bv_r} \ket{\bu_r}\ket{\lambda_r}
    \left[ \sqrt{1-(\frac{c}{\lambda_r})^2}\ket{0} + \frac{c}{\lambda_r}\ket{1}   \right],
    \label{eq:sigma_inv}
\end{equation}
where $c$ is a constant to ensure the inverse of eigenvalues are no larger than 1. 

We can repeatedly perform a conditional measurement of \eqref{eq:sigma_inv} until it is in state $\ket{1}$. After that, we can uncompute and discard the eigenvalue register which results in 
\begin{equation}
    \ket{\psi_1} \equiv \frac{1}{\sqrt{p(1)}} \sum_{r=1}^{R} \frac{c}{\sigma_{r}} \ket{\bv_r} \ket{\bu_r},
    \label{eq:psi_1}
\end{equation}
where $p(1) = \sum_{r=1}^{R} |\frac{c}{\lambda_r}|^2$ is the probability of the qubit in \eqref{eq:sigma_inv} collapsing to state $\ket{1}$.

\vspace*{0.1in}

\noindent{\sc{Step 3: Calculating the inner products.}}

Follow the notations in \eqref{eq:psi_X} --\eqref{eq:y_x}, we can rewrite \eqref{eq:yhat_svd} as 
\begin{equation}
    \sum_{r=1}^{R}\sigma_r^{-1} \braket{\widetilde{\bx} | \bv_r} \braket{\by | \bu_r}.
    \label{eq:inner}
\end{equation}
Motivated by the strategy in \cite{schuld2016prediction}, we write \eqref{eq:inner} into some entries of a qubit so that it can be assessed by a single measurement. To this end, we construct two states $\ket{\psi_1}$ and $\ket{\psi_2}$ that are entangled with a qubit, i.e.
\begin{equation}
    \frac{1}{\sqrt{2}} \left( \ket{\psi_1}\ket{0} + \ket{\psi_2}\ket{1} \right),
    \nonumber
\end{equation}
where $\ket{\psi_1}$ is defined in \eqref{eq:psi_1} and $\ket{\psi_2}=\ket{{\by}}\ket{{\widetilde{\bx}}}$. Then, the off-diagonal elements of this qubit's density matrix read
\begin{equation*}
    \frac{c}{2\sqrt{p(1)}} \sum_{r=1}^{R}\sigma_r^{-1} \braket{\widetilde{\bx} | \bv_r} \braket{\by | \bu_r} = \frac{c}{2\sqrt{p(1)}} \sum_{r=1}^{R} \sigma_{r}^{-1} \widetilde{\bx}^{\T} \bv_r 
    \bu_r^{\T} {\by},
    \label{eq:inner_2}
\end{equation*}
which contain the desired results \eqref{eq:yhat_svd} up to a known normalization factor.

Given a testing set of $n_t$ observations, i.e. $\{\widetilde{y}_i, \widetilde{\bx}_i\}_{i=1}^{n_t}$, the prediction error can be calculated as 
\begin{equation}
    \tilde{g}(\ket{i}) = \frac{1}{n_t} \sum\limits_{i=1}^{n_t} (\widehat{y}_i - \widetilde{y}_i )^2,
    \label{eq:pred_error}
\end{equation}
where $\ket{i}$ is a quantum state represents $\{ x_1, \ \ldots \ , x_d \}$ and $\widehat{y}_i$ is the predictor of  $\widetilde{y}_i$ which can be calculated follow the QLP  procedure introduced above. In addition, the computational complexity of running QLP over a training set of size $n$ and a testing set of $n_t$ is approximately of the order $O\left((n+n_t)\log_2 d\right)$ which is faster than the classical computing algorithm that is usually of order $O\left((n+n_t)d\right)$.

\section{Hybrid quantum  adaptive  search}
\label{sec:implement}

\begin{wrapfigure}[9]{r}{0.5\textwidth} 
  \vspace{-25pt}
  \begin{center}
    \includegraphics[width=0.5\textwidth]{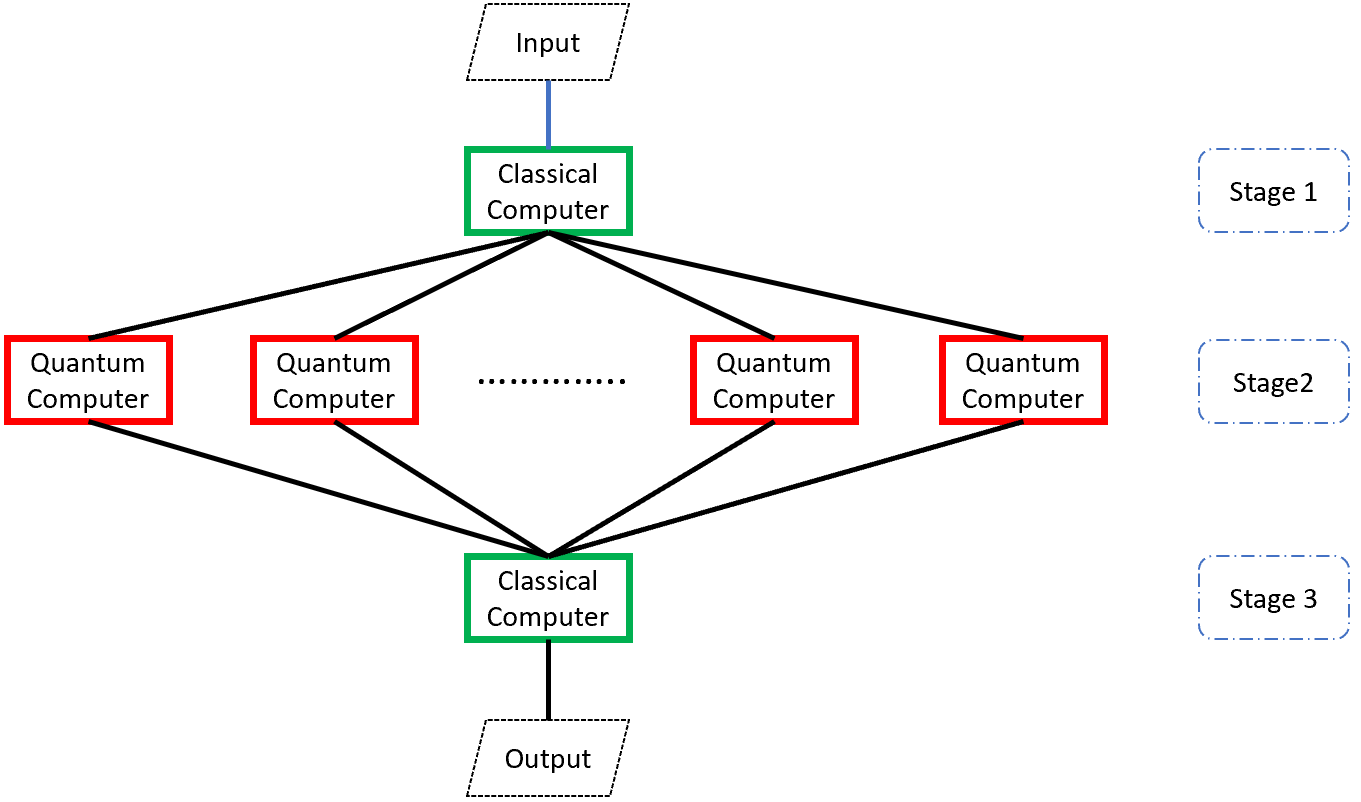}
    \captionsetup{width=.5\textwidth}
    \caption{Flowchart of hybrid quantum-classical strategy.}
    \label{fig:flowchart}
  \end{center}
\end{wrapfigure}

In the high dimensional regime, implementing the best subset selection procedure completely on a quantum computing system is desirable since both QAS and QLP are substantially faster than their classical counterparts.
However, such an objective may not be readily accomplished due to the prototypical development of quantum computers. On one hand, the capacity of the state-of-the-art quantum processor (about 70 qubits) is still far from large enough to carry out big data applications. For example, the quantum state preparation step of QLP requires at least $2(\log_2 d + \log_2 n)$ qubits which may exceed the capacity of most public available quantum computers when $d$ and $n$ are moderate or large. On the other hand, existing quantum computing systems are usually highly sensitive to the environment and can be influenced by both internal and external noises \citep{sung2019non}. For example, a superconducting quantum computing system can be affected by the internal noises due to material impurities as well as the external noises that come from control electronics or stray magnetic fields \citep{Kandala_2019}.
Further, such noises can accumulate through the computing process and create a non-negligible bias.

To address the aforementioned practical challenges, we propose a hybrid quantum-classical strategy to balance computational efficiency, scalability, and reliability of quantum best subset selection.
The intuition of the hybrid quantum-classical strategy is to advance complementary advantages of quantum and classical computing by implementing computational demanding steps on quantum nodes and running capacity demanding steps on classical nodes. The accuracy of quantum nodes can be boosted by some aggregation approaches, like majority voting. 
To be specific, we first input the data into a classical node and parallelly compute the linear prediction errors \eqref{eq:pred_error} over  $D=2^p$ candidate models. The linear prediction errors, stored in a $D$-dimensional vector, will be passed to $K$ quantum nodes. Then, we independently implement QAS on $K$ quantum nodes to select the minimum element in the $D$-dimensional prediction error vector. The selection results, stored in a list of $K$ items, will be passed to a classical node for a majority voting. The best subset with the most votes is selected as the final estimator. We present the flow chart and details of this procedure in Figure \ref{fig:flowchart} and  Algorithm \ref{algo:hybrid}, respectively.

\begin{algorithm}
    \caption{Hybrid quantum-classical strategy for best subset selection}
    \label{algo:hybrid}
    \begin{algorithmic}
    \State \textbf{Input}: A training set $\{\bx_i, \ y_i\}_{i=1}^n$ with $\bx_i \in \RR^{p}$. A testing set $\{\widetilde{\bx}_i\}_{i=1}^{n_t}$ with $\widetilde{\bx}_i \in \RR^{p}$. A number of available quantum nodes $K$. A learning rate $\lambda \in (0, 1)$.
    
        \State \textbf{Stage 1 on a classical computer:}
        \State \quad (1.1) Compute, in parallel, the prediction error \eqref{eq:pred_error} over all $D=2^p$ candidate models.
        \State \quad (1.2) Save the prediction errors $\bG_D= \left(g(\ket{0}), \ \ldots, \  g(\ket{D-1})\right)$.
        
        \State \textbf{Stage 2 on $K$ quantum nodes:}
        \State \quad (2.1) Independently implement Algorithm \ref{algo:qas} over $\bG_D$ on $K$ quantum nodes.
        \State \quad (2.2) Save the results in a list of $K$ selected models $\cM = \{m_1, \ \ldots, \  m_K\}$.
        
        \State \textbf{Stage 3 on a classical computer:}
        \State \quad\quad\quad A majority voting over models in $\cM$. Denote the model with the most votes as $\widehat{m}$.

        \State \textbf{Output}: The selected model $\widehat{m}$.
    \end{algorithmic}
\end{algorithm}

The Theorem \ref{thm:vote} below states a majority voting over $K=2\xi+1$ independent quantum nodes can boost the success probability of finding the best subset. The lower bound of success probability in Theorem \ref{thm:vote} can be boosted by increasing the number of quantum nodes or improving the success probability on each quantum node.

\begin{theorem}\label{thm:vote}
Let's consider a majority voting system that consists of $K=2\xi+1$ nodes, where $\xi$ is a positive integer. Suppose that each node works independently with a probability $q>\frac{\xi+1}{2\xi+1}$ to vote the correct model and a probability $1-q$ to vote an incorrect model. Let $\cal E$ denote the event that the majority voting system selects the correct model. Then, the probability of $\cal E$ is lower bounded by
\begin{align*}
    \mathbb{P}({\cal E}) > \Phi\left( \sqrt{2(2\xi+1)D_{KL}(q,\frac{\xi+1}{2\xi+1})} \right), 
\end{align*}
where $\Phi(\cdot)$ is the cumulative distribution function of a standard normal random variable and
$$D_{KL}(a,b)=b\ln\frac{b}{a}+(1-b)\ln\frac{(1-b)}{(1-a)}$$
is the Kullback-Leibler divergence between two Bernoulli distributions with parameters $a$ and $b$, respectively.
\end{theorem}

\begin{wrapfigure}[7]{r}{0.5\textwidth} 
  \vspace{-55pt}
  \begin{center}
    \includegraphics[width=0.45\textwidth]{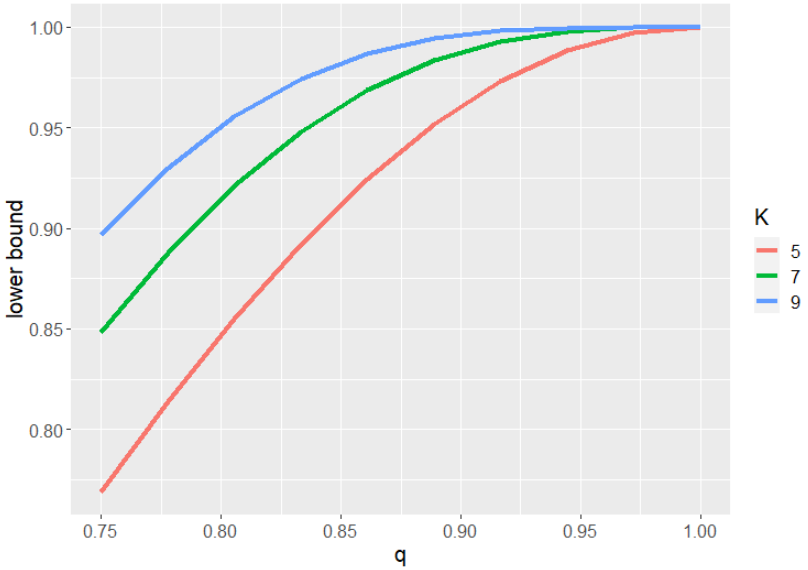}
    \vspace{-5pt}
    \captionsetup{width=.5\textwidth}
    \caption{Lower bounds of successful probability when $K = 5,7,9$.}
    \label{fig:lower bound}
  \end{center}
\end{wrapfigure}

In Figure \ref{fig:lower bound}, we plot the lower bound of $\mathbb{P}({\cal E})$ versus the values of $q$. We let $q$ increase from 0.75 to 1 and choose the number of nodes $K = 5,7,9$. According to Figure \ref{fig:lower bound}, all three solid lines are concave curves which means a majority voting system with a moderate number of nodes can effectively boost the success probability of finding the correct model.  For instance, a majority voting over  $K=9$ independent nodes with $q=0.75$ can boost the probability of finding the correct model to 0.9 or higher.




\section{Experiments on the quantum computer}
\label{sec:exp_quantum}

In this section, we implement the proposed hybrid quantum-classical strategy on IBM Quantum Experience. This platform has developed a Qiskit Python development kit\footnote{\url{https://qiskit.org/}}, which allows users to perform both quantum computing and classical computing in a single project. In Section \ref{sec:exp_tuning}, we analyze the performance of Algorithm \ref{algo:hybrid} regarding the selections of tuning parameters. In Section \ref{sec:exp_QAS}, we assess the best subset section performance of Algorithm \ref{algo:hybrid} under various linear regression settings and compare QAS with Grover's algorithm in stage 2.

\subsection{Selection of tuning parameters}
\label{sec:exp_tuning}

The proposed hybrid quantum-classical strategy involves two tuning parameters: the number of quantum nodes $K$; and a learning rate $\lambda \in (0,1)$ which will be passed to each quantum node to implement Algorithm \ref{algo:adaptive_search}.  As suggested by Theorem \ref{thm:vote}, the success probability of Algorithm \ref{algo:hybrid} converges to 1 as $K$ increases when the success probability of each quantum node surpasses 0.5. Therefore, one should choose $K$ as large as possible subject to the availability of quantum computing resources. On the other hand, the learning rate $\lambda$ controls the ``step size" of each iteration in Algorithm \ref{algo:qas}. In this subsection, we use an experiment to assess the performance and the sensitivity of Algorithm \ref{algo:hybrid} with respect to the choices of $K$ and $\lambda$.
To focus on the selection of tuning parameters, we skip the stage 1 in Algorithm \ref{algo:hybrid}. Instead, we generate $\bG_D$ as an i.i.d. sample of size $D=32$ from the uniform distribution $U[0,1]$. Then we use the stages 2 and 3 in Algorithm \ref{algo:hybrid} to find the minimum element in $\bG_D$. We set $K=1$, $3$ or $5$, and let $\lambda$ be a sequence of grid points between 0.40 and 0.60 with a step size of 0.01.  For each pair of $K$ and $\lambda$, we simulate 200 replications. The performance is measured by the accuracy rate which is defined as
\begin{equation}
    \text{Accuracy rate} = \frac{\text{Number of replications with correct solution}}{\text{Number of replications}}.
    \label{eq:ar}
\end{equation}

\begin{wrapfigure}[8]{r}{0.5\textwidth} 
  \vspace{-15pt}
  \begin{center}
    \includegraphics[width=0.5\textwidth]{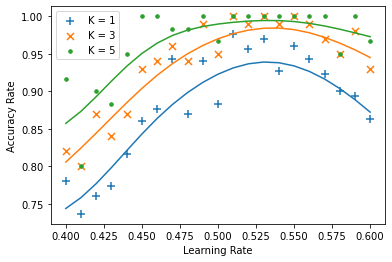}%
    \vspace{-7pt}
    \captionsetup{width=.5\textwidth}
    \caption{Accuracy rates of Algorithm \ref{algo:hybrid} with $K=1,3,5$ and $\lambda \in [0.40, 0.60]$. }
    \label{fig:accuracy_rate}
  \end{center}
\end{wrapfigure}

The results are presented in Figure \ref{fig:accuracy_rate}. The accuracy rates for $K=1, 3$ or $5$ are plotted as blue, orange, and green dots with different symbols. The solid lines are corresponding smoothed curves. According to Figure \ref{fig:accuracy_rate}, we find that increasing the number of quantum nodes $K$ can improve the accuracy rate though the improvement is marginal when $K$ is large enough, e.g. $K=5$. Besides, we observe that the highest accuracy rates are attained with some $\lambda$ between 0.5 and 0.55 for both choices of $K$. One can notice that the accuracy rates are close to 1 when $\lambda \in (0.5, 0.55)$ and $K \geq 3$ which can be considered as a rule-of-thumb recommendation. Further, the smoothed curves in Figure \ref{fig:accuracy_rate} indicate that Algorithm \ref{algo:hybrid} is not very sensitive for the selection of tuning parameters.

\subsection{Best subset selection with hybrid quantum-classical strategy}\label{sec:exp_QAS}

In this subsection, we assess the best subset selection performance of the proposed hybrid quantum-classical strategy. We consider a linear regression model
\begin{equation}
    y_i = \bx_i^{\T}\bbeta^{\ast} + \epsilon_i, \quad i=1, \ \ldots, \  n,
    \label{eq:sim_linear}
\end{equation}
where $y_i \in \RR$, $\bx_i \in \RR^p$, $\bbeta^{\ast} = (\beta_1, \ \ldots \ , \beta_p)^{\T} \in \RR^p$, and $\epsilon_i \in \RR$. First, we draw $\{\bx_i\}_{i=1}^n$ as an i.i.d. sample from a multivariate normal distribution $N_p({\bf 0}, \bSigma)$, where the $(i,j)$th entry of $\bSigma$ equals $\rho^{|i-j|}$ for some $\rho \in (0,1)$. Then, we draw $\{\epsilon_i\}_{i=1}^n$ as an i.i.d. sample from a normal distribution $N(0, \sigma^2)$. Notice that, $\rho$ controls the autocorrelation level among covariates, and $\bbeta^{\ast}$, $\rho$ and $\sigma$ together control the signal to noise ratio (SNR), i.e. $\text{SNR} = \bb^{\ast\T} \bSigma \bb^{\ast} / \sigma^2$. Let $s<p$ be a positive integer indicating the model sparsity. We set the first $s$ elements of $\bbeta^{\ast}$ to be 1 and the rest $p-s$ elements to be 0.

Subject to the capacity of the quantum computing system, we set $n = 100$, $p =7$, and $s=4$. We choose the autocorrelation level $\rho \in \{0.25,  0.5\}$ and  the signal to noise ratio $\text{SNR} \in \{0.5, 1.0, 2.0, 3.0\}$, respectively. For each scenario, we simulate 200 replications. We implement the hybrid quantum-classical strategy as proposed in Algorithm \ref{algo:hybrid} with $K=3$ and $\lambda=0.55$ to select the best subset of the linear regression model. We denote this method as {\sc QAS} since it implements the quantum adaptive search in stage 2.
Besides, we consider two variants of Algorithm \ref{algo:hybrid} as two competitors. The first competitor, denoted as {\sc Grover Oracle}, replaces {\sc QAS} in stage 2 by Grover's algorithm with a known oracle state. The second competitor, denoted as {\sc Grover Random}, replaces {\sc QAS} in stage 2 by Grover's algorithm with a randomly selected oracle state. Notice that {\sc Grover Oracle} is an oracular method and not applicable in practice since the oracle state is not observable. 

For each replication, we record the false positives ({\sc FP}) and false negatives ({\sc FN}) of each method. {\sc FP} is the number of inactive covariates that are selected into the model and {\sc FN} is the number of active covariates that are ignored in the selected model. The box-plots of {\sc FP} and {\sc FN} over 200 replications are reported in Figure \ref{fig:ibm} below. According to Figure \ref{fig:ibm}, {\sc Grover Oracle} performs the best among the three competitors which is not surprising as it utilizes the oracle information of the true best subset. The proposed {\sc QAS} method, as a non-oracular approach, performs almost identical to {\sc Grover Oracle} in nearly all scenarios. {\sc QAS} is slightly outperformed by {\sc Grover Oracle} when the correlation level is high ($\rho=0.5$) and the signal to noise ratio is small ($\text{SNR}=0.5$), which is the most challenging scenario. In contrast, {\sc Grover Random} performs as bad as random guesses in all scenarios. This observation, which is inline with the discussions in Section \ref{sec:limitation}, shows oracular quantum search algorithms are not suitable to statistical learning problems.

\begin{figure}[H]
  \begin{center}
    \includegraphics[scale=1]{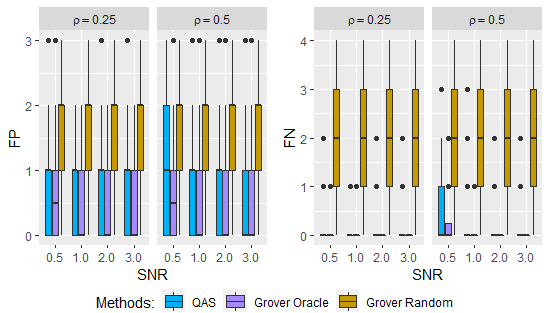}
    \vspace{-5pt}
    \caption{Box-plots of false positives (left) and false negatives (right) over 200 replications.}  
  \label{fig:ibm}
  \end{center}
\end{figure}


\section{Comparison with classical methods}
\label{sec:sim_classical}

In this section, we compare the empirical performance of quantum adaptive search ({\sc QAS}) with the naive best subset selection method ({\sc BSS}) implemented on a classical computer. We  also include some popular approximate best subset selection methods designed for classical computers: {\sc forward stepwise} selection \citep[e.g.][]{draper1966applied}; {\sc LASSO}  \citep{tibshirani1996regression}; and the fast best subset selection method ({\sc L0Learn}) \citep{hazimeh2018fast}. The implementation details of the 5 comparison methods are summarized in Table \ref{tab:sim_imp}. 

\begin{table}[ht]
\centering
\caption{Implementation details for the 5 comparison methods in Section \ref{sec:sim_classical}.}
\label{tab:sim_imp}
\vspace{-10pt}
\begin{tabular}{@{}llllc@{}}
\toprule
Method name &  & Computing method  &  & Tuning parameter selection  \\ \midrule
{\sc QAS }  &  & Algorithm \ref{algo:hybrid}  &  & $K=5$ and $\lambda=0.5$ \\
{\sc BSS}  &  & Naive best subset selection &  & NA \\
{\sc Forward stepwise} &  & R package bestsubset\footnote{Avaliable at \url{https://github.com/ryantibs/best-subset}}       &  & NA  \\ 
{\sc LASSO}   &  & R package glmnet\footnote{Avaliable at \url{https://cran.r-project.org/web/packages/glmnet}}   &  & 10-folds cross-validation                    \\
{\sc L0Learn}        &  & R package L0Learn\footnote{Avaliable at \url{https://cran.r-project.org/web/packages/L0Learn}} &  & 10-folds cross-validation  \\ \bottomrule
\end{tabular}
\end{table}

We consider the same linear regression model in \eqref{eq:sim_linear}. The coefficient vector $\bbeta^{\ast}$ is generated from one of the following two settings.
\begin{itemize}
    \item[(1)] {\textbf{ Strong sparsity:}} the first $s$ elements of $\beta^*$ equal to 1 while the rest $p-s$ elements equal to 0.
    \item[(2)] {\textbf{ Weak sparsity:}} the first $s$ elements of $\beta^*$ equal to $1, \frac{s-1}{s}, \frac{s-2}{s}, \ \ldots, \  \frac{1}{s}$ while the rest $p-s$ elements equal to 0.
\end{itemize}
In this experiment, we set $n=100$, $p=10$ and $s=5$.
We choose the autocorrelation level $\rho \in \{0.25,  0.5\}$ and  the signal to noise ratio $\text{SNR} \in \{0.5, 1.0, 2.0, 3.0\}$, respectively. For each scenario, we simulate 100 replications. Additional results for $p=20$ and a real data analysis are relegated to the supplemental material. Besides the false positives ({\sc FP}) and false negatives ({\sc FN}) defined in Section \ref{sec:exp_QAS}, we measure the prediction performance of each method by the relative test error ({\sc RTE}): 
\begin{equation*}
    \text{RTE}=\mathbb{E}(\widetilde{y} - \widetilde{\bx}^T\widehat{\bbeta})^2/\sigma^2 = (\widehat{\bbeta}-\bbeta^\ast)^{\T} \bSigma (\widehat{\bbeta}-\bbeta^\ast)/ \sigma^2 + 1,
\end{equation*}
where $(\tilde{y}, \tilde{\bx})$ is a testing observation which is i.i.d. with the training sample $\{y_i, \bx_i\}_{i=1}^n$. It is easy to see that {\sc RTE} is lower bounded by 1 and the smaller the better.

\begin{figure}[H]
  \begin{center}
    \includegraphics[scale=0.54]{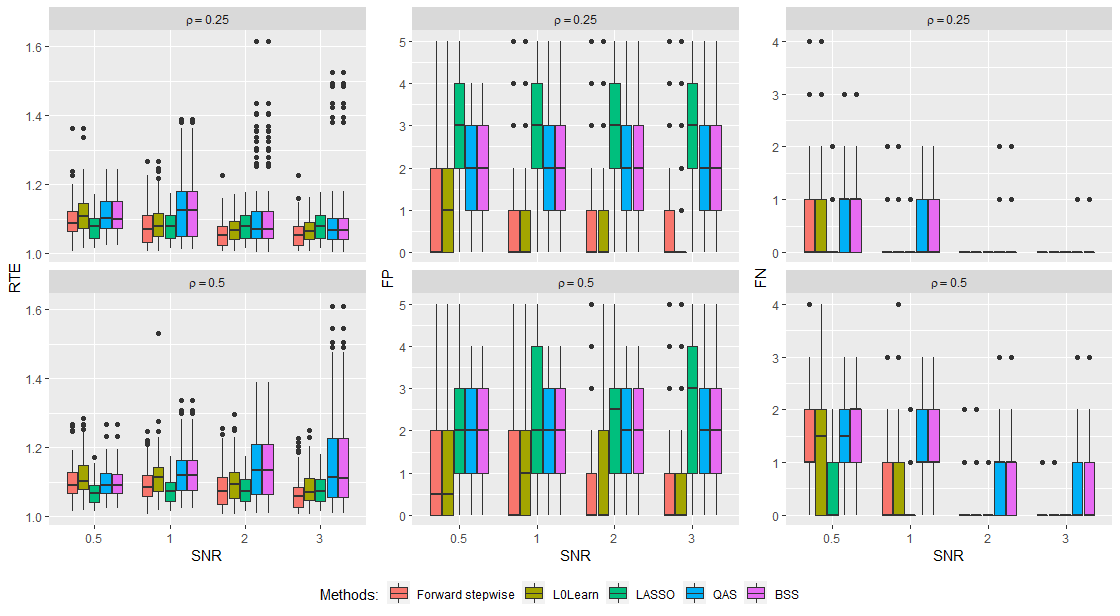}
    \vspace{-3pt}
    \caption{{\bf Strong sparsity setting:} Boxplots of relative test error (left), false positives (middle) and false negatives (right) over 100 replications.}   
  \label{fig:strong_01}
  \end{center}
\end{figure}

\begin{figure}[H]
  \begin{center}
    \includegraphics[scale=0.54]{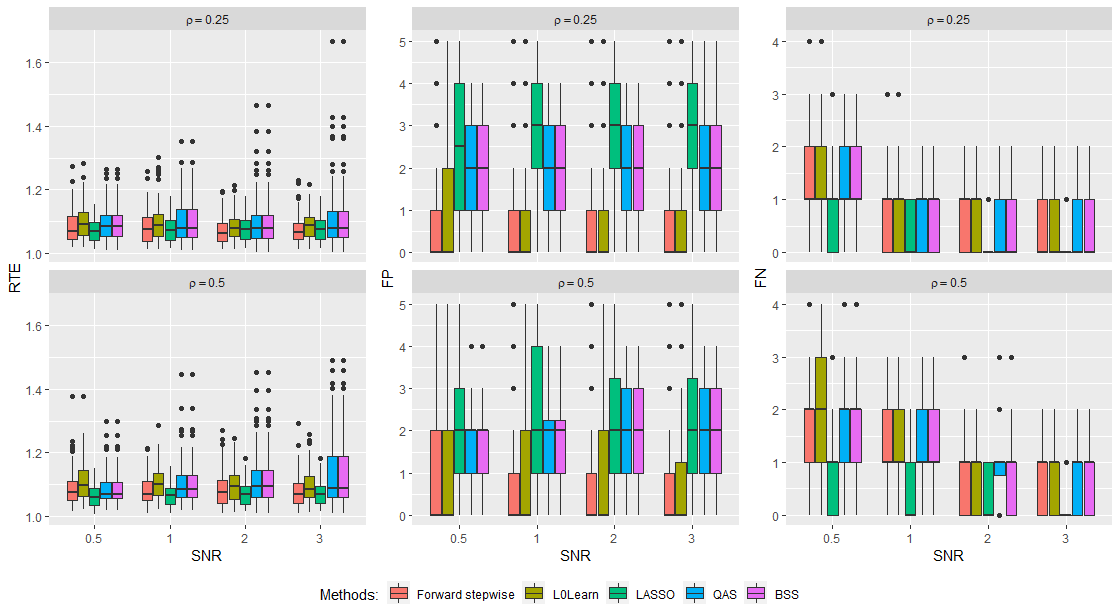}
    \vspace{-3pt}
    \caption{{\bf Weak sparsity setting:} Boxplots of relative test error (left), false positives (middle) and false negatives (right) over 100 replications.}   
  \label{fig:weak_01}
  \end{center}
\end{figure}

Figures \ref{fig:strong_01} and \ref{fig:weak_01} present box-plots of {\sc FP}, {\sc FN} and {\sc RTE} over 100 replications under strong sparsity and weak sparsity settings, respectively. According to the box-plots, none of the 5 competing methods dominate the others in terms of all three measurements, which is inline with the discussions in \cite{hastie2017extended}. Generally speaking, {\sc LASSO} tends to select 
generous models which may have small {\sc FN}s but large {\sc FP}s. {\sc Forward stepwise} and {\sc L0Learn} prefer to select parsimonious models which may lead to small {\sc FP}s but large {\sc FN}s. However, none of the three methods can be considered as a good approximation of {\sc BSS} since their performance are distinct in most scenarios. In contrast, {\sc QAS} performs almost identical to {\sc BSS} in terms of all three measurements in every scenario.

\begin{figure}[H]
  \begin{center}
    \includegraphics[scale=0.48]{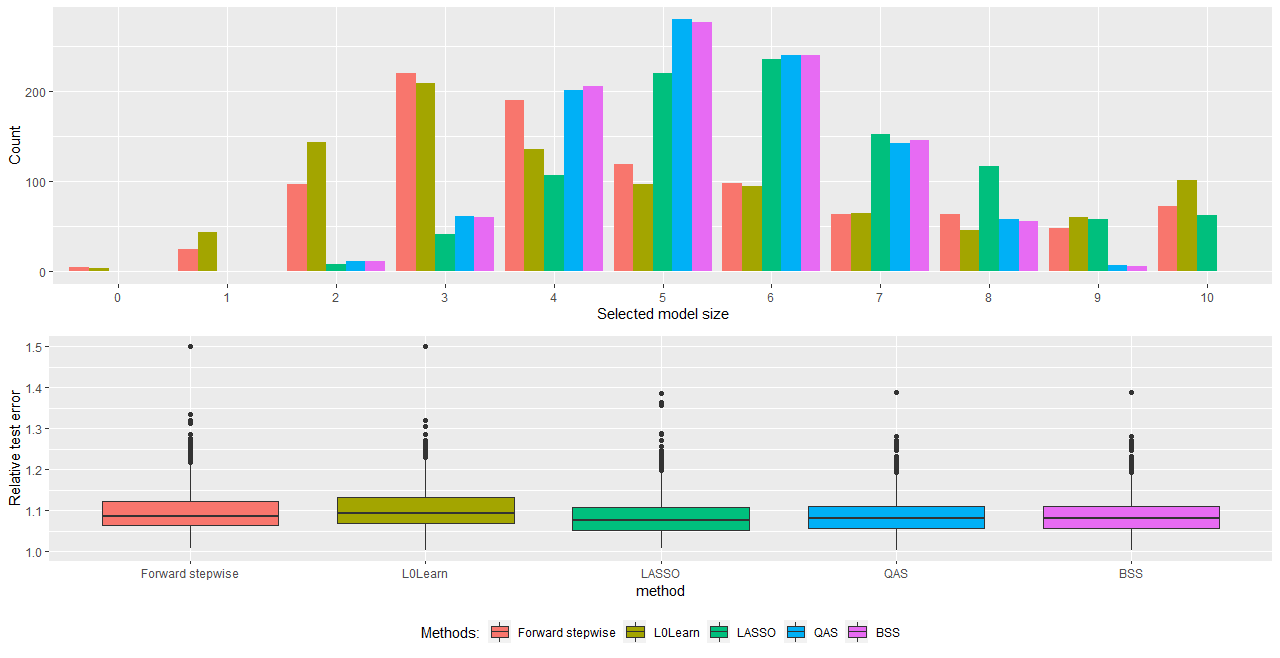}
    \vspace{-5pt}
    \caption{{\bf Best subset selection behaviors:} histogram of selected model size (top) and    boxplots of relative test error (bottom).}   
  \label{fig:boot}
  \end{center}
\end{figure}

Next, we take a closer look at best subset selection behaviors of the 5 comparison methods. We simulate 1,000 replications of \eqref{eq:sim_linear} with $n=100$, $p=10$, $s=5$, $\rho=0.5$, $\text{SNR}=0.5$, and the strong sparsity $\bbeta^*$. In the top panel of Figure \ref{fig:boot}, we present the histogram of selected model sizes for 5 methods. In the bottom panel of Figure \ref{fig:boot}, we report the box-plots of {\sc RTS}s for 5 methods. According to Figure \ref{fig:boot}, the selected model sizes of {\sc QAS} and {\sc BSS} sharply concentrates around the truth, i.e. $s=5$. The histograms of {\sc Forward stepwise} and {\sc L0Learn} are severely left skewed which indicates they tend to underestimate the size of the active set. In contrast, {\sc LASSO} has a right skewed histogram and hence often select oversized models. Again, the selection as well as prediction performances of {\sc QAS} and {\sc BSS} are almost identical. The results in Figure \ref{fig:boot} further justified our argument that {\sc QAS} is an efficient quantum alternative of {\sc BSS} while {\sc Forward stepwise},  {\sc LASSO} and {\sc L0Learn} are not ideal approximates of {\sc BSS}.

\section{Conclusion}
\label{sec:conclusion}

In this paper, we investigated the solution of the best subset selection problem in a quantum computing system. We proposed an adaptive quantum search algorithm that does not require the oracle information of the solution state and can significantly reduce the computational complexity of classical best subset selection methods. An efficient quantum linear prediction method was also discussed. Further, we introduced a novel hybrid quantum-classical strategy to avoid the hardware bottleneck of existing quantum computing systems and boost the success probability of quantum adaptive search. The quantum adaptive search is applicable to various best subset selection problems. Moreover, our study opens the door to a new research area: reinvigorating statistical learning with powerful quantum computers.

\newpage
\bibliographystyle{apalike}
\bibliography{reference/QAS}
\end{document}


\title{Supplementary Material of ``Best Subset Selection:  Statistical Computing Meets Quantum Computing"}

\author{Wenxuan Zhong, Yuan Ke, Ye Wang, Yongkai Chen, Jinyang Chen, Ping Ma
\thanks{Department of Statistics,  University of Georgia, Athens, GA 30602. E-mail:
\href{mailto:wenxuan@uga.edu}{\textsf{wenxuan@uga.edu, yuan.ke@uga.edu, ywang927@uga.edu, YongkaiChen@uga.edu, jingyang.chen@uga.edu, pingma@uga.edu}}.}}

\date{}
\maketitle
\vspace{-0.5in}

This supplement provides technical proofs and additional numerical results. In Section \ref{app:theorem}, we present the proofs of
theoretical results in Sections 3 and 4. In Section
\ref{sec:add_num}, we report additional simulation results and a real data
application to complement the numerical analysis in the main document.

\setcounter{equation}{0}
\setcounter{section}{0}
\setcounter{subsection}{0}

\baselineskip=24pt

\renewcommand{\theequation}{S.\arabic{equation}}
\renewcommand{\thetable}{S\arabic{table}}
\renewcommand{\thefigure}{S\arabic{figure}}
\renewcommand{\thesection}{S.\arabic{section}}
\renewcommand{\thelemma}{S.\arabic{lemma}}

\appendix
\section{Proofs}
\label{app:theorem}

\subsection{Technical Lemmas}

In this subsection, we provide two technical lemmas to pave the way for the proof of main theorems. We omit the proof of Lemma \ref{lem:a1} as it can be found in the literature.

\begin{lemma}[c.f. Lemma 1 in \cite{boyer1998tight}]\label{lem:a1}
For any real numbers $\alpha$ and $\beta$ and any positive integer $m$, we have
\begin{equation}
    \sum\limits_{j=0}^{m-1} \cos(\alpha+2\beta j) = \frac{\sin(m\beta)\cos(\alpha+(m-1)\beta)}{\sin \beta}. \nonumber
\end{equation}
In particular, when $\alpha=\beta$, the above equality can be simplified as
\begin{equation}
    \sum\limits_{j=0}^{m-1} \cos(\alpha+2\beta j) = \frac{\sin(2m\alpha)}{2\sin{\alpha}}.
    \nonumber
\end{equation}
\end{lemma}

\begin{lemma}\label{lem:a2}
Let $s_0$ be the (unknown) number of solutions states among $D$ states. Let $\theta$ be such that $\sin^2\theta = s_0/D$. Let $\gamma$ be an arbitrary positive integer. Let $j$ be an integer sampled uniformly between 0 and $\tau-1$. If we observe the register after applying $j$ Grover's operations to the initial state $\ket{\psi_0}=\sum_{i=0}^{D-1} \frac{1}{\sqrt{D}}\ket{i}$, the probability of obtaining a solution is exactly
\begin{equation}\label{lem:a1}
    P_{\tau} = \frac{1}{2} - \frac{\sin(4\tau\theta)}{4\tau\sin(2\theta)}.
\end{equation}
Further, we have $P_{\tau} \geq \frac{1}{4}$ when ${\tau} \geq \frac{1}{\sin(2\theta)}$.
\end{lemma}

\begin{proof}[Proof of Lemma \ref{lem:a2}]
The probability of finding one solution state among $s_0$ if we perform $j$ Grover's operations is $s_0k_j^2 =\sin^2((2j+1)\theta)$. It follows
that the average success probability when $0 \leq  j < \tau$ is chosen randomly is
\begin{align*}
    P_\tau & = \sum_{j=0}^{\tau-1}\frac{1}{\tau}\sin^2((2j+1)\theta)
    \\
    & = \frac{1}{2\tau} \sum_{j=0}^{\tau-1} \{1-\cos((2j+1)2\theta)\}
    \\
    & = \frac{1}{2} - \frac{\sin(4\tau\theta)}{4\tau\sin{2\theta}}.
\end{align*}

If $\tau\geq \frac{1}{\sin(2\theta)}$, we  complete the proof as the following inequality holds
\begin{equation*}
    \frac{\sin(4\tau\theta)}{4\tau\sin{2\theta}}\leq \frac{1}{4\tau\sin{2\theta}}\leq \frac{1}{4}.
\end{equation*}

\end{proof}

\subsection{Proof of Theorem 3.1}

For the ease of presentation and without loss of generality, we re-numbering the states in descending order in this proof, i.e.
\begin{equation}
    g(\ket{0}) >   g(\ket{1}) \geq g(\ket{2}) \geq \ \cdots \ \geq g(\ket{D-1}) > g(\ket{D}),
    \label{eq:g_rank_des}
\end{equation}
where $\ket{D}$ is the unique solution state and $\ket{0}$ is an added initial state to facilitate the discussion with $g(\ket{0})$ being an arbitrarily large value.

Let's assume Algorithm 2 is initialized with the least favorable state $\ket{0}$. Let $Z$ denote the number of iterations the algorithm takes to arrive at the solution state $\ket{D}$. The rule that Algorithm 2 moves from $\ket{0}$ to $\ket{D}$ can be abstracted as the following mathematical process.
\begin{itemize}
    \item[]{\sc Iteration 1}: Draw an integer $X_1$ uniformly from $0$ to $D$, then the algorithm moves from $\ket{0}$ to $\ket{X_1}$.
    \item[]{\sc Iteration 2}: Draw an integer $X_2$ uniformly from $0$ to $D-X_1$, then the algorithm moves from $\ket{X_1}$ to $\ket{X_1+X_2}$.
    \item[] \begin{center} $\vdots$   \end{center}
    \item[]{\sc Iteration $z$}: Draw an integer $X_z$ uniformly from $0$ to $D-\sum_{i=1}^{z-1}X_i$, then the algorithm moves from $\ket{\sum_{i=1}^{z-1}X_i}$ to $\ket{\sum_{i=1}^{z}X_i}$.
    If $\sum_{i=1}^zX_i = D$, then the algorithms stops  at $Z=z$ as it finds the solution state $\ket{D}$. Otherwise, the algorithm goes to the $(z+1)$th iteration.
\end{itemize}

As we can see, the total number of iterations $Z$ is a discrete random variable that can take any positive integer values. To prove Theorem 1, it is equivalent to show that the $\kappa$th quantile of the discrete random variable $Z$ is upper bounded by $C_\kappa\ln D$ for a positive constant $C_\kappa$. The proof will be unveiled in three steps.

\medskip
\noindent{\bf Step 1: A partial sum process.}

To investigate the probability distribution of $Z$, we first study a partial sum of $X_i$ which is defined as follows
\begin{align*}
    S_z = \sum_{i = 1}^z X_i \;\; \text{for} \;\; z = 1,2, \ \ldots.
\end{align*}

Notice that $(S_{z}|S_{z-1}=s_{z-1}, \ \ldots, \  S_{1} = s_1) \stackrel{\mathcal{D}}{=} (S_{z}|S_{z-1}=s_{z-1}) \sim \text{unif}\{s_{z-1}, D\}$. Also, we can show $\{S_z\}_{z=1, 2, \ \ldots}$ is a submartingale, i.e. $\mathbb{E}[S_{z}|S_1, \ \ldots \ , S_{z-1}] = \mathbb{E}[S_{z}|S_{z-1}] \geq S_{z-1}$. Thus, $\{S_z\}_{z = 1, 2,\  \ldots}$ is a discrete-time Markov chain with a finite state space $\{0,1,2,\ \ldots \ , D\}$.

Moreover, the expectation of $S_z$ satisfies
\begin{align*}
     \mathbb{E}[S_z] & = \mathbb{E}\left[\mathbb{E}[S_z | S_{z-1}]\right] = \mathbb{E}\left[\frac{S_{z-1}+D+1}{2}\right] 
    \\
    &= \mathbb{E}[S_{z-1}]/2 + (D+1)/2
    \\
    & = \mathbb{E}[X_1]/2^{z-1} + (D+1)\big(1 - 1/2^{z-1}\big) 
    \\
    &= (D+1)/2^{z} + + (D+1)\big(1 - 1/2^{z-1}\big)
    \\
    & = (D+1)\big(1- 1/2^z\big).
\end{align*}

\medskip
\noindent{\bf Step 2: The probability mass function of $Z$.}

Next, we derive the probability mass function of $Z$. The derivation is based on the Markov property of the partial sum process $\{S_z\}_{z = 1, 2, \ \ldots}$. Define a transition matrix $P$ as
\begin{align*}
  P = \left( \begin{array}{cccccc}
  \frac{1}{D+1} &  \frac{1}{D+1}  &  \frac{1}{D+1}  &  \cdots  &  \frac{1}{D+1}  &  \frac{1}{D+1}  \\
  0 &   \frac{1}{D} & \frac{1}{D} &  \cdots          &  \frac{1}{D}   &  \frac{1}{D}  \\
  0 &   0   &  \frac{1}{D-1}  & \cdots &  \frac{1}{D-1}  &  \frac{1}{D-1}  \\
  \vdots &   \vdots   & \vdots &  \ddots  & \vdots &  \vdots  \\
  0 &   0   &  0  & \cdots &  \frac{1}{2}  &  \frac{1}{2}  \\
  0 &   0   &  0  &  \cdots  & 0 &  1  \end{array} \right).
\end{align*}
Let $\pi_z$ be a row vector with dimension $D+1$ that denotes the distribution of the chain $\{S_z\}_{z=1,2,\ldots}$ at time $z$. By the definition of $\{S_z\}_{z=1,2,\ \ldots}$ and $P$, we have
\begin{align*}
    \pi_{z+1} = \pi_z P, \quad \text{for} \quad z = 1, 2, \ \ldots .
\end{align*}
Hence, we can write out $\pi_1$, $\pi_2$ and $\pi_3$ as
\begin{align*}
    \pi_1 = \Big(\frac{1}{D+1}, \frac{1}{D+1}, \frac{1}{D+1}, \ \ldots \ , \frac{1}{D+1}, \frac{1}{D+1} \Big),
\end{align*}
\begin{align*}
    \pi_2 = \frac{1}{D+1}\bigg(\frac{1}{D+1}, \sum_{i_1=D}^{D+1}\frac{1}{i_1}, \sum_{i_1=D-1}^{D+1}\frac{1}{i_1}, \ \ldots \ , \sum_{i_1=2}^{D+1}\frac{1}{i_1}, \sum_{i_1=1}^{D+1}\frac{1}{i_1} \bigg), \quad \text{and}
\end{align*}
\begin{align*}
    \pi_3 = \frac{1}{D+1}\bigg(\frac{1}{(D+1)^2}, \sum_{i_2=D}^{D+1}\frac{1}{i_2}\sum_{i_1=i_2}^{D+1}\frac{1}{i_1}, \sum_{i_2=D-1}^{D+1}\frac{1}{i_2}\sum_{i_1=i_2}^{D+1}\frac{1}{i_1}, \ \ldots \ , \sum_{i_2=2}^{D+1}\frac{1}{i_2}\sum_{i_1=i_2}^{D+1}\frac{1}{i_1}, \sum_{i_2=1}^{D+1}\frac{1}{i_2}\sum_{i_1=i_2}^{D+1}\frac{1}{i_1} \bigg).
\end{align*}

In general, we can summarize the expression of $\pi_z$ for $z = 1, 2, \ \ldots$ as
\begin{align*}
    \pi_z = \Big(\mathbb{P}(S_z = 0), \mathbb{P}(S_z = 1), \mathbb{P}(S_z = 2), \ \ldots \ , \mathbb{P}(S_z = D-1),\mathbb{P}(S_z = D)  \Big),
\end{align*}
where
\begin{align*}
    \mathbb{P}(S_z = j) = \frac{1}{D+1}\sum_{i_{z-1}=D+1-j}^{D+1}\frac{1}{i_{z-1}}\sum_{i_{z-2}=i_{z-1}}^{D+1}\frac{1}{i_{z-2}}\ \cdots\ \sum_{i_1=i_2}^{D+1}\frac{1}{i_1},\quad \text{for} \quad j = 0,1, \ \ldots \ , D.
\end{align*}

Therefore, we can write out the probability mass function of $Z=z$ for $z = 1,2,\ \dots$ as
\begin{align}
    \mathbb{P}(Z = z) = \mathbb{P}\big(S_z = D \big) = \frac{1}{D+1}\sum_{i_{z-1}=1}^{D+1}\frac{1}{i_{z-1}}\sum_{i_{z-2}=i_{z-1}}^{D+1}\frac{1}{i_{z-2}} \ \cdots \ \sum_{i_2=i_3}^{D+1}\frac{1}{i_2}\sum_{i_1=i_2}^{D+1}\frac{1}{i_1}.
    \label{eq:pmf_Z}
\end{align}

Notice that the last summation in \eqref{eq:pmf_Z}, i.e. $\sum_{i_1=i_2}^{D+1}\frac{1}{i_1}$ , 
 is  a finite partial sum of a harmonic series. Therefore, we can write it as
\begin{align*}
    \sum_{i_1=i_2}^{D+1}\frac{1}{i_1} = \ln (D+1) + \eta_{D+1} - \ln (i_2) - \eta_{i_2} \asymp \ln (D+1) - \ln (i_2),
\end{align*}
where $\eta_{z} \asymp \frac{1}{2z}$. 

Let us move on to the next level of summation in \eqref{eq:pmf_Z}, which is upper bounded by 
\begin{align*}
    \sum_{i_2=i_3}^{D+1}\frac{1}{i_2}\sum_{i_1=i_2}^{D+1}\frac{1}{i_1} &\asymp \sum_{i_2=i_3}^{D+1}\frac{1}{i_2}\big[\ln (D+1) - \ln (i_2)\big] \lesssim \ln(D+1)\big[ \ln(D+1) - \ln(i_3) \big] \lesssim \ln^2(D+1),
\end{align*}
and lower bounded by
\begin{align*}
    \sum_{i_2=i_3}^{D+1}\frac{1}{i_2}\sum_{i_1=i_2}^{D+1}\frac{1}{i_1} &\asymp \sum_{i_2=i_3}^{D+1}\frac{1}{i_2}\big[\ln (D+1) - \ln (i_2)\big] \asymp \ln(D+1)\big[ \ln(D+1) - \ln(i_3) \big] - \sum_{i_2 = i_3}^{D+1}\frac{\ln(i_2)}{i_2}\\
    &\gtrsim \ln^2(D+1) - \ln(D+1)\ln(i_3),
\end{align*}
since $\sum_{i_2 = i_3}^{D+1}\frac{\ln(i_2)}{i_2} \asymp \frac{1}{2}\ln^2(D+1) - \frac{1}{2}\ln^2(i_3)$.

Similarly, we can go through all the summations in \eqref{eq:pmf_Z} and show
\begin{align}
    \mathbb{P}(Z = 0) = \frac{1}{D+1}  \quad \text{and} \quad \mathbb{P}(Z = z) \asymp \frac{\ln^{z-1}(D+1)}{D+1}.
    \label{eq:pmf_Z_2}
\end{align}

\medskip
\noindent{\bf Step 3: The $\kappa$th quantile of $Z$.}

With the results in \eqref{eq:pmf_Z_2}, we summarize the cumulative distribution function of $Z$ as 
\begin{align*}
    F_Z(z) = \mathbb{P}(Z \leq z) = \sum_{j = 0}^z\mathbb{P}(Z = j) \asymp \frac{1}{D+1}+\sum_{j=1}^{z}\frac{\ln^{j-1}(D+1)}{D+1} \asymp \frac{\ln^{z-1}(D+1)}{D+1}.
\end{align*}

Let $z = c\ln(D+1)+1$ with a positive constant  $c \in (0,1)$, we have
\begin{align*}
    F_Z(z) \asymp \frac{\ln^{z-1}(D+1)}{D+1} \asymp \frac{(D+1)^{c\ln\ln(D+1)}}{D+1} \asymp 1.
\end{align*}
Therefore, for any $\kappa \in (\frac{1}{2}, 1)$, there exists a positive constant $C_\kappa$ such that the $\kappa$th quantile of $Z$ is upper bounded by $C_\kappa\log_2{D}$ which completes the proof.

\qed

\subsection{Proof of Theorem 3.2}\label{sec:proof_Thm32}

Suppose we are in the $m$th iteration of Algorithm 2. Let $\ket{w_{m}}$ be the current benchmark state and $r_m$ be the rank of $g(\ket{w_{m}})$ in the sorted  sequence of $\{g{\ket{i}}\}_{i=0}^{D-1}$ in ascending order, $r_m = 1,\ldots ,D$. 
Notice that, Algorithm 2 implements $\tau(m)=\lceil \pi \lambda^{-m/2} /4\rceil\equiv \lceil \frac{\pi}{4}\gamma^m\rceil$ Grover's operations, where $\gamma=\lambda^{-1/2}$. 
We are interested in finding the expected number of Grover's operations to update $\ket{w_{m}}$.

Let $\theta$ be the angle such that $\sin^2 \theta=r_m/D$. Let
\begin{align*}
    \tau^*_m= \frac{1}{\sin(2\theta)}=\frac{D}{2\sqrt{(D-r_m)r_m}}<\sqrt{\frac{D}{r_m}}.
\end{align*}
We say that Algorithm 2 reaches a phase transition for $\ket{w_{m}}$ if $\tau(s)$ exceeds $\tau^* (m)$ for some $s\geq m$.

The expected total number of Grover's operations needed to reach the phase transition for $\ket{w_{m}}$ is upper bounded by
\begin{align*}
    \frac{\pi}{4}\sum_{j=0}^{\lceil \log_\gamma (\tau^*_m) \rceil} \gamma^{j} 
    < \frac{\pi}{4} \frac{\gamma^m-1}{\gamma-1}<\frac{\tau^*_m+1}{\gamma-1}.
\end{align*}
Thus, if the algorithm updates $\ket{w_{m}}$ before it reaches reach the phase transition, the expected number of Grover's operations is at most of the order $O(\tau^*_m)$, which is further upper bounded by $O(\sqrt{{D}/{r_m}})$.

If the  phase  transition for $\ket{w_{m}}$ is reached, as proved by Lemma \ref{lem:a2},  every new iteration of Algorithm 2 will be able to update $\ket{w_{m}}$ with a probability at least 1/4 since $\tau(s)\geq \tau^*_m$. Then, the expected iterations to update $\ket{w_{m}}$  after the phase transition is upper bounded by a positive constant. Hence, the expected number of additional Grover's operations needed to update $\ket{w_{m}}$ 
after the phase transition is upper bounded by $C\tau^*_m$ for some positive constant $C$. These two scenarios together complete the proof.

\qed

\subsection{Proof of Theorem 4.1}

Recall that the $K=2\xi+1$ nodes work independently and share the same ``success" probability $q$. Let $B \equiv B(\xi, q)$ denote the random variable that indicates the number of nodes who vote for the correct model in the system. It is easy to see that $B$ follows a binomial distribution with parameters $2\xi+1$ and $q$.

Then, the probability of $\cal E$ follows
\begin{align}
    {\mathbb P({\cal E})} & = \sum_{i=\xi+1}^{2\xi+1} p_B(i)= 1-\sum_{i=0}^{\xi} p_B(i)= 1- F_B(\xi), \nonumber
\end{align}
where $p_B(i)$ and $F_B(a)$ are the probability mass function and the cumulative distribution function (CDF) of $B$, respectively.

Theorem 1 in \cite{short2013improved} provides an upper bound for $F_B(\xi)$, i.e., 
\begin{align}
    F_B(\xi) < \Phi\left(\mathrm{sign}(\xi+1-Kq) \sqrt{2KD_{KL}(q,\frac{\xi+1}{K})} \right),
    \nonumber
\end{align}
where $\Phi(\cdot)$ is the CDF of a standard normal random variable and $\mathrm{sign}(\cdot)$ is the sign function. Further
$$D_{KL}(a,b)=b\ln\frac{b}{a}+(1-b)\ln\frac{(1-b)}{(1-a)}$$
is the Kullback-Leibler divergence between two Bernoulli distributions with parameters $a$ and $b$, respectively.

Thus, we  bound the probability of $\cal E$ from below,
\begin{align*}
    {\mathbb P({\cal E})} & > 1 - \Phi\left(\mathrm{sign}(\xi+1-Kq) \sqrt{2KD_{KL}(q,\frac{\xi+1}{K})} \right)
    \\
    & = \Phi\left(\mathrm{sign}(Kq-\xi-1) \sqrt{2KD_{KL}(q,\frac{\xi+1}{K})} \right)
    \\
    & = \Phi\left( \sqrt{2(2\xi+1)D_{KL}(q,\frac{\xi+1}{2\xi+1})} \right),
\end{align*}
where the last equality uses the fact $K=2\xi+1$ and the condition $q>\frac{\xi+1}{2\xi+1}$.
\qed

\section{Additional numerical results}
\label{sec:add_num}
\subsection{Additional results for Section 6}

\begin{figure}[H]
  \begin{center}
    \includegraphics[scale=0.48]{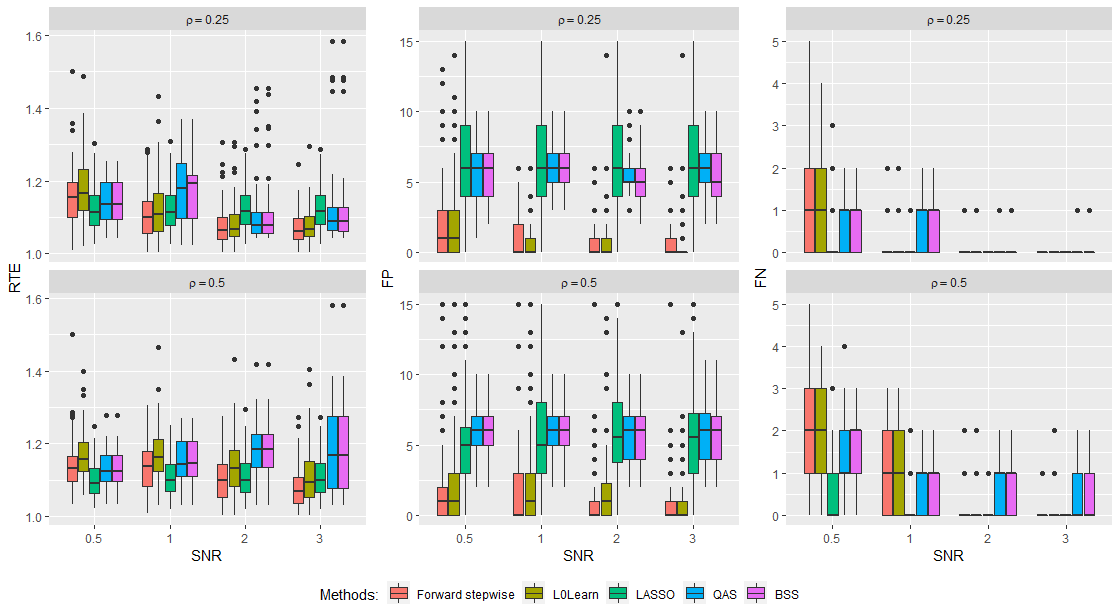}
    \vspace{-10pt}
    \caption{{\bf Strong sparsity setting with $p=20$:} Boxplots of relative test error (left), false positives (middle) and false negatives (right) over 100 replications.}  
  \label{fig:strong_02}
  \end{center}
\end{figure}

\begin{figure}[H]
  \begin{center}
    \includegraphics[scale=0.48]{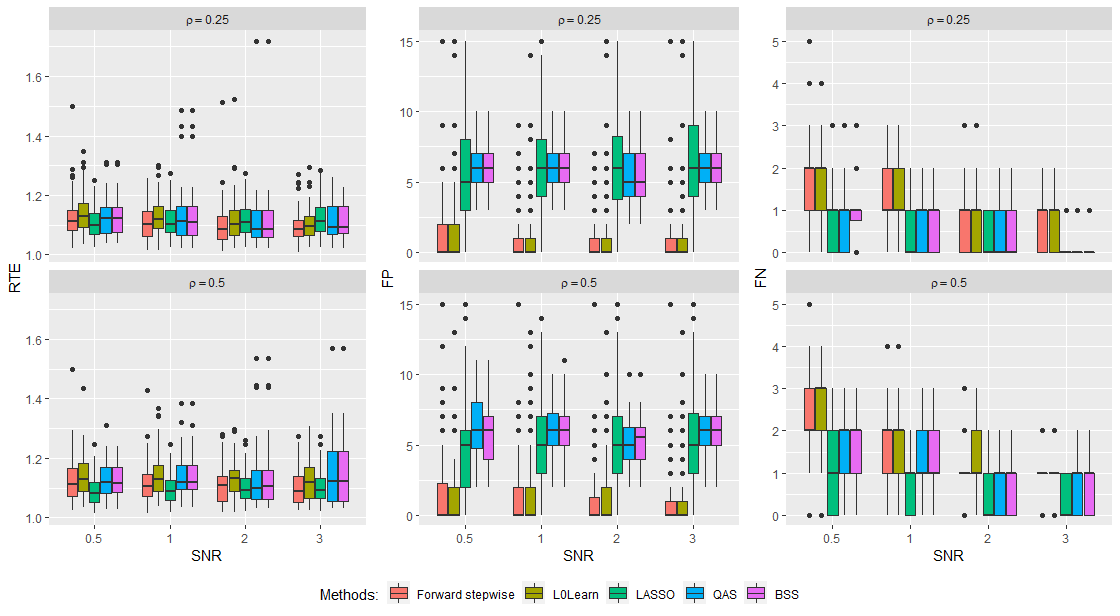}
    \vspace{-10pt}
   \caption{{\bf Weak sparsity setting with $p=20$:} Boxplots of relative test error (left), false positives (middle) and false negatives (right) over 100 replications.}   
  \label{fig:weak_02}
  \end{center}
\end{figure}

In this subsection, we report additional simulation results for Section 6. All the simulation settings are the same as introduced in Section 6 except that we increase $p$ to 20. Figures \ref{fig:strong_02} and \ref{fig:weak_02} present box-plots of {\sc FP}, {\sc FN} and {\sc RTE} over 100 replications under strong sparsity and weak sparsity settings, respectively. The results for $p=20$ are similar as the results we reported in the main document for $p=10$.

\vspace{-10pt}

\subsection{Real data analysis}

We compare the performance of the methods listed in Table 1 on a real dataset. The dataset \citep{johnson1996fitting} contains 18 measurements including body fat, age, weight, height, and ten body circumference measurements, e.g., abdomen, on 252 men. The percentage of body fat (\textit{brozek}) is considered as the response. More details are summarized in Table \ref{tbl:fat_data_descirption}.

\begin{table}[H]
\small
\centering
\caption{Descriptions of 18 health measurements in the dataset \citep{johnson1996fitting}.}
\label{tbl:fat_data_descirption}
\begin{tabular}{c|c}
\toprule\hline
\textbf{variable} & \textbf{description}                                                                \\ \hline
\textit{brozek}            & Percent body fat using Brozek's equation                                            \\ \hline
\textit{density}           & Density (gm/$cm^3$)                                                                 \\ \hline
\textit{weight}            & Weight(lbs)                                                                         \\ \hline
\textit{adipos}            & Adiposity index = Weight/Height$^2$ (kg/$m^2$)                                      \\ \hline
\textit{neck}              & Neck circumference (cm)                                                             \\ \hline
\textit{abdom}            & Abdomen circumference (cm) at the umbilicus and level with the iliac crest          \\ \hline
\textit{thigh}            & Thigh circumference (cm)                                                            \\ \hline
\textit{ankle}             & Ankle circumference (cm)                                                            \\ \hline
\textit{forearm}          & Forearm circumference (cm)                                                                                                  \\ \hline
\textit{siri}              & Percent body fat using Siri's equation                                              \\ \hline
\textit{age}              & Age(yrs)                                                                            \\ \hline
\textit{height}           & Height(inches)                                                                      \\ \hline
\textit{free}              & Fat Free Weight = (1 - fraction of body fat) * Weight, using Brozek's formula (lbs) \\ \hline
chest             & Chest circumference (cm)                                                            \\ \hline
\textit{hip}               & Hip circumference (cm)                                                              \\ \hline
\textit{knee}              & Knee circumference (cm)                                                             \\ \hline
\textit{biceps}           & Extended biceps circumference (cm)                                                  \\ \hline
\textit{wrist}             & Wrist circumference (cm) distal to the styloid processes                            \\ \hline\bottomrule
\end{tabular}
\end{table}

The goal is to find a linear model that can accurately predict body fat (\textit{brozek}) using the other variables except for \textit{siri} (another way of computing body fat), \textit{density} (it is used in the brozek and siri formulas) and \textit{free} (it is computed using brozek formula). So the total number of predictors is 14. We randomly split the dataset into 80\% training and 20\% testing for 100 replications.  The out-of-sample mean-squared error (Test MSE) is used as an evaluation metric in this study. In addition, we report the support size, i.e., the number of non-zero coefficients.

\begin{figure}[H]
  \begin{center}
    \includegraphics[scale=0.48]{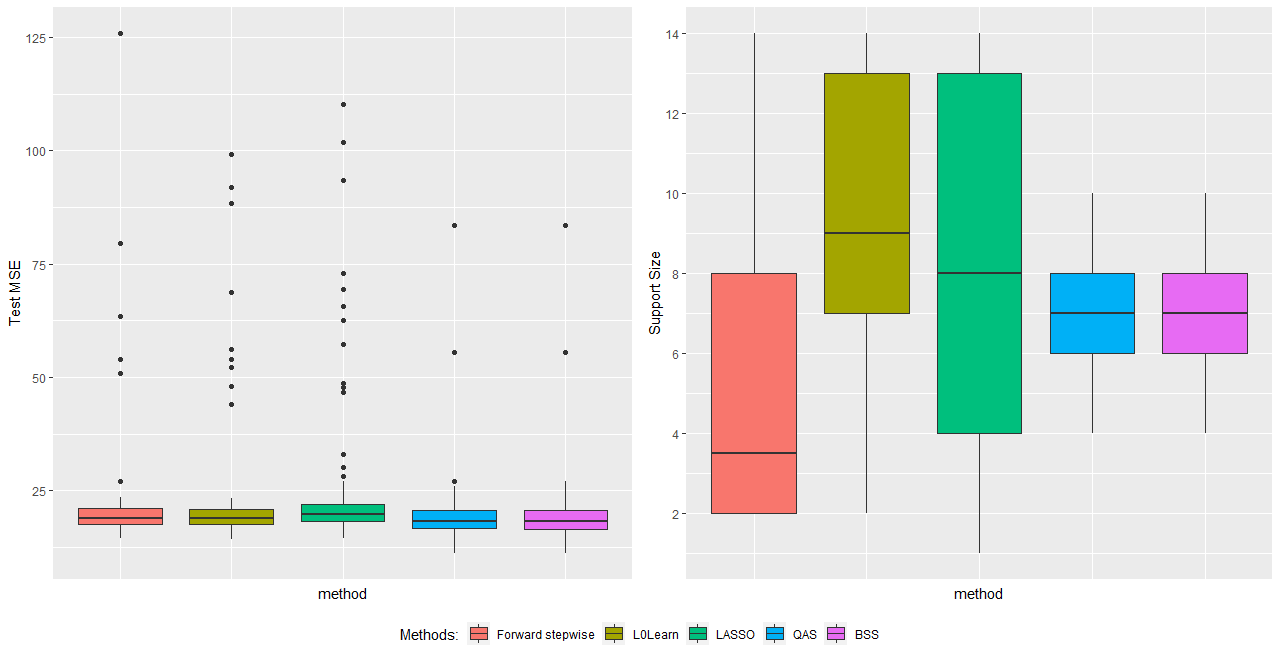}
    \vspace{-5pt}
    \caption{Left: boxplots of Test MSE for five methods. Right: boxplots of support sizes for five methods.  }   
  \label{fig:realdata}
  \end{center}
\end{figure}

\begin{table}[ht]
\centering
\caption{Test MSEs and support sizes for five methods on the body fat dataset.}
\label{tab:realdata}
\begin{tabular}{@{}lcccccc@{}}
\toprule
&        & \multicolumn{1}{l}{Forward stepwise} & \multicolumn{1}{l}{L0Learn} & \multicolumn{1}{l}{LASSO} & \multicolumn{1}{l}{QAS} & \multicolumn{1}{l}{BSS} \\ \midrule
Test MSE & median & 18.92 & 18.85 & 19.81 & \textbf{18.12} & \textbf{18.12} \\
& mean   & 21.85 & 23.00  & 25.87 & 19.52 & \textbf{19.51} \\
& sd     & 13.85 & 15.24  & 18.04 & \textbf{8.06} & \textbf{8.06} \\ \midrule
Support Size & median & 3.50 & 9.00 & 8.00 & \textbf{7.00} & \textbf{7.00} \\
& mean   & \textbf{5.17} & 9.49 & 8.21 & 6.99 & 6.99 \\
& sd     & 3.43 & 3.44 & 4.36 & \textbf{1.31} & \textbf{1.31} \\ \bottomrule
\end{tabular}
\end{table}

According to Figure \ref{fig:realdata} and Table \ref{tab:realdata}, we observe that QAS and BSS perform identical and outperform the other methods. 
Moreover, Table  \ref{tbl:realdata_selected_counts} shows the percentage of each predictor that has been selected within $100$ repetitions across five different methods. According to Table \ref{tbl:realdata_selected_counts}, we can see \textit{abdom} is the key factor in predicting the percentage of body fat. This finding is inline with the studies in \cite{ochiai2010} and \cite{flegal2009}.

\begin{table}[H]
\centering
\caption{Measurements selected percentage}
\label{tbl:realdata_selected_counts}
\begin{tabular}{@{}cccccc@{}}
\toprule
Rank & Forward Stepwise & L0Learn       & LASSO         & QAS           & BSS           \\ \midrule
1    & abdom(100\%)         & abdom(100\%)      & abdom(100\%)      & abdom(100\%)      & abdom(100\%)      \\
2    & weight(74\%)     & wrist(92\%)   & wrist(98\%)   & wrist(92\%)   & wrist(92\%)   \\
3    & wrist(67\%)      & age(78\%)     & age(97\%)     & weight(84\%)  & weight(84\%)  \\
4    & neck(38\%)       & height(73\%)  & height(94\%)  & forearm(84\%) & forearm(84\%) \\
5    & forearm(36\%)    & forearm(71\%) & forearm(77\%) & neck(74\%)    & neck(74\%)    \\
6    & age(33\%)        & neck(60\%)     & neck(76\%)    & thigh(67\%)   & thigh(68\%)   \\
7    & biceps(28\%)     & biceps(52\%)  & ankle(66\%)   & age(62\%)     & age(61\%)     \\
8    & height(22\%)     & thigh(49\%)   & biceps(64\%)  & hip(54\%)     & hip(54\%)     \\
9    & hip(22\%)        & knee(45\%)    & thigh(52\%)   & biceps(26\%)  & biceps(26\%)  \\
10   & thigh(22\%)      & ankle(45\%)   & hip(49\%)     & height(18\%)  & height(18\%)  \\
11   & adipos(21\%)     & hip(44\%)     & weight(48\%)  & chest(12\%)   & chest(12\%)   \\
12   & chest(20\%)       & weight(41\%)  & knee(48\%)    & adipos(11\%)  & adipos(11\%)  \\
13   & ankle(18\%)      & chest(40\%)    & adipos(40\%)   & knee(11\%)    & knee(11\%)    \\
14   & knee(16\%)       & adipos(32\%)  & chest(40\%)    & ankle(4\%)   & ankle(4\%)   \\ \bottomrule
\end{tabular}
\end{table}

\bibliographystyle{apalike}
\bibliography{reference/QAS}